\documentclass[times, twoside]{arxivStyle}
\usepackage{pict2e}
\usepackage{picture}
\usepackage{blindtext}

\usepackage[color=blue!20]{todonotes}
\usepackage{listings}
\usepackage{multicol}
\usepackage{multirow}
\usepackage{pdflscape}
\usepackage[inkscapelatex=false]{svg}
\usepackage{xr}
\usepackage{makecell}
\usepackage{adjustbox}
\usepackage{diagbox}
\usepackage{array}
\usepackage{tikz}
\usepackage[scaled=0.9]{inconsolata}
\usepackage{enumitem}
\usepackage[commandnameprefix=always]{changes}
\usepackage{nameref}

\usepackage{textcomp,mathcomp,amsmath}

\makeatletter
\newcommand{\namedsection}[2]{%
  \expandafter\gdef\csname sectionname@#1\endcsname{#2}%
  \label{#1}%
}
\newcommand{\secnameref}[1]{\csname sectionname@#1\endcsname}
\makeatother

\leadauthor{Albrecht, Schmidt} 

\begin{document}

% -------------------------------
% --- frontmatter: Title page ---
% -------------------------------
\title{\huge Album: executable building blocks for scientific imaging routines, from sharing to LLM-assisted orchestration}
% \title{\huge Album: executable building blocks for scientific imaging routines, from sharing solutions to automated conversion of research scripts into replicable software} -lang, aber akkurat ;) 
\shorttitle{Album}
%\letter
\author[1,2,3,$\ast$]{Jan Philipp Albrecht}
\author[1,2,$\ast$]{Deborah Schmidt}
\author[1,2]{Lucas Rieckert}
\author[1,2]{Maximilian Otto}
%\author[1,2]{ABC}
%\author[1,2]{ABC}
\author[1,4,\Letter]{Kyle Harrington}

\affil[1]{Max Delbr\"uck Center for Molecular Medicine in the Helmholtz Association, Berlin, Germany}
\affil[2]{
Helmholtz Imaging, Max Delbr\"uck Center for Molecular Medicine in the Helmholtz Association, Berlin, Germany}
\affil[3]{Humboldt-Universität zu Berlin, Faculty of Mathematics and Natural Sciences, Berlin, Germany}
\affil[4]{Biohub, Redwood City, CA, USA}
\affil[$\ast$]{equal contributions} 

\maketitle

\begin{abstract}
Open-source scientific software is a major driver of scientific progress, yet its development and reuse remain difficult in collaborative settings. Researchers repeatedly face four recurring challenges: discovering and reproducing existing routines, adapting them for new use cases, sharing and scaling them across collaborators, and stabilizing them with reproducible execution environments. We present Album, an open-source framework for packaging and sharing scientific routines as executable artifacts through two minimal primitives: (i) the solution, a Python-native executable entry point that combines machine-readable metadata, arguments, environment specifications, and lifecycle hooks; and (ii) the catalog, a decentralized, git-native distribution mechanism with indexed search and optional web rendering for discovery, provenance, and governance. Album uses a two-context execution model in which a host controller evaluates manifests and prepares per-solution environments, while lifecycle hooks execute inside isolated solution environments. This design supports reproducible execution, post-environment setup, and the composition of routines with incompatible dependencies. Album can be used in conjunction with LLM agents: solutions can be drafted and revised with LLM assistance, and a proof-of-concept MCP interface exposes cataloged solutions as callable tools for tool-grounded discovery and orchestration. We evaluate Album through four real-world imaging deployments spanning interactive visualization of electron microscopy data, integration of multiple segmentation methods, the orchestration
of cryo-electron tomography competition workflows, and mineral quantification pipelines. Overall, Album complements package managers, workflow systems, and container runtimes by making scientific routines executable, shareable artifacts. Documentation and examples are available at \url{https://album.solutions/}.
\end{abstract}


\begin{figure*}[tp!]
    \centering
    \includegraphics[width=\textwidth]{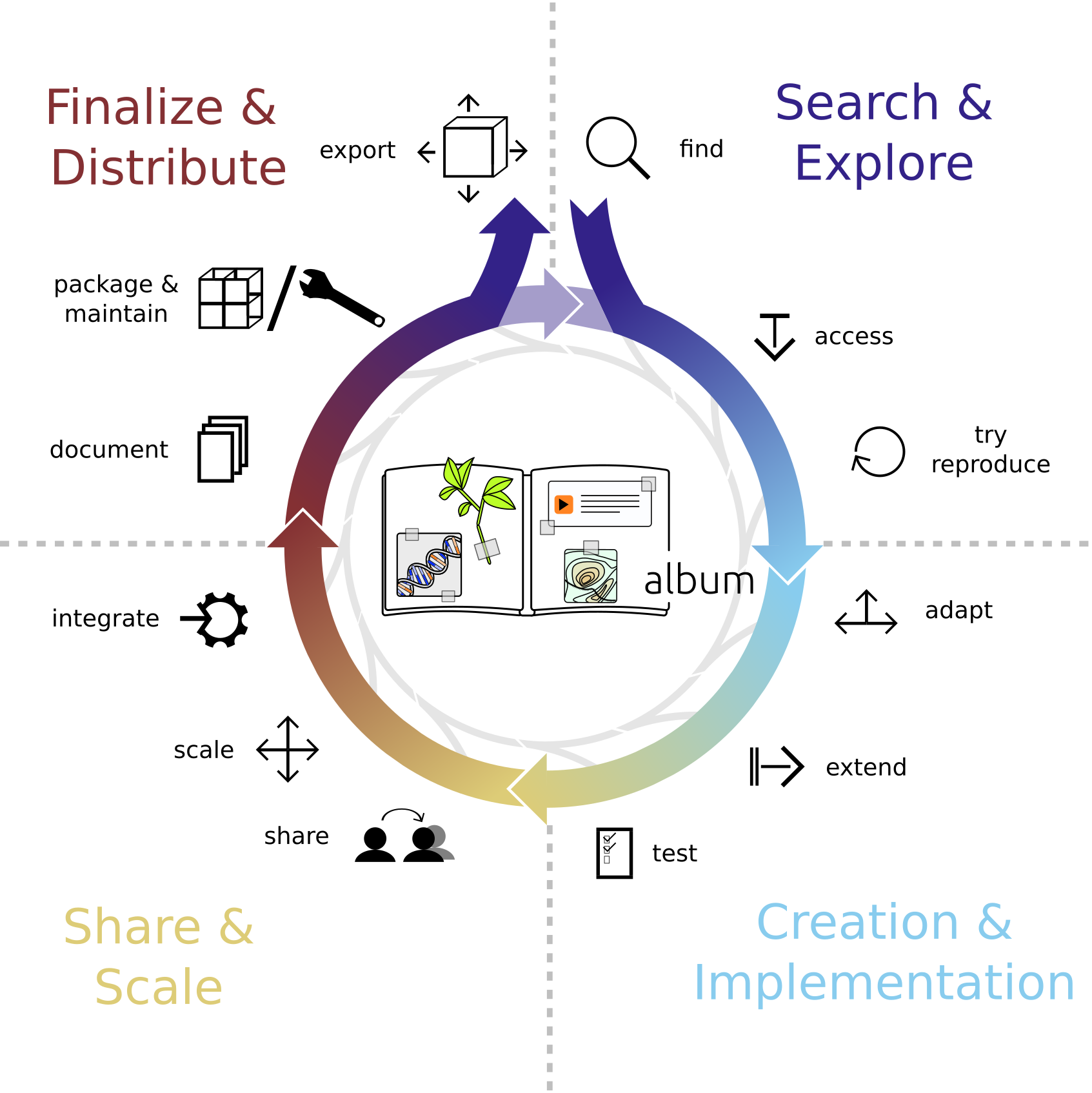}
   \caption{\textbf {Iterative, collaboration-driven research software cycle}. The cycle consists of four stages: Search and Explore (top right), Creation and Implementation(bottom right), Share and Scale (bottom left) and Finalize and Distribute (top left). In Search and Explore tasks are related to findability, accessibility and reproducibility of a scientific work. Creation and Implementation is all about adapting or extending the work to the new use case whereas Share and Scale involves tasks such as sharing and scaling the work but might also contain the task of integrating the work into a larger context. Finalize and Distribute usually comes last and involves tasks such as documentation and packaging. All stages come with their own challenges that are addressed by our framework.}
    \label{fig:icdssdp}
\end{figure*}

\section*{Introduction}
%%% pargraph 1: introduce collaboration in science in biomedical research
Research software and research methods have become indispensable for modern science, enabling data analysis, visualization, and exploratory workflows across domains \cite{hocquet_software_2024}. In parallel, contemporary research increasingly relies on collaboration, within and across laboratories and across disciplines, as reflected by rising co-authorship and cross-field cooperation \cite{wachter_karrieren_2023, mazzocchi_scientific_2019}. Yet, despite the central role of software and collaboration, the development, sharing, and maintenance of research routines remain difficult, particularly when collaborators have heterogeneous technical expertise and when interactive tools are integral to the scientific process.
%%% paragraph 2: what is the nature of the challenge? technical problems with workflows and many components for differnet components (of the software development cycle)
Scientific software users often span a spectrum from \emph{true users}, who rely on accessible graphical interfaces, to \emph{power users}, who extend and compose methods by building on prior work \cite{cohen-boulakia_search_2011}. Effective collaboration between these groups is crucial for project success \cite{tobias_fourteen_2023, segal_software_2009}, but it is frequently impeded by practical barriers: locating and reusing existing routines, adapting them to a new scientific context, sharing intermediate results for rapid feedback, and distributing robust, reproducible versions for broader reuse. These challenges are amplified in domains such as bioimage analysis, where many domain experts lack formal computational training \cite{cimini_crucial_2024}.
%%% paragraph 3: pinpoint problems
Existing distribution and reproducibility technologies address parts of this problem. Package managers and archive networks improve reuse, while containers provide strong isolation and replication guarantees \cite{spinellis_package_2012, kurtzer_singularity_2017, moreau_containers_2023}. However, these approaches often impose substantial overhead on users and collaborators, and they are frequently misaligned with the realities of iterative scientific development. Research software typically evolves through incremental cycles \cite{boehm_spiral_1988, ruparelia_software_2010, djaffardjy_developing_2023}, requiring frequent execution by collaborators. In practice, collaborators may lack permissions or expertise to reproduce environments, and crucial dependencies are not always available through standard package channels. Moreover, many scientific routines require \emph{post-environment} steps such as downloading external assets, compiling components, enabling GUI-based interactivity, or integrating OS-specific tooling (e.g., Blender-based visualization pipelines). These steps are difficult to express and share consistently using conventional environment files alone, and containerization can further complicate interactive execution.
%%% paragraph 4: ponpoint our approach to the problem above
In this article, we present first present an \emph{Iterative, collaboration-driven research software cycle} that captures recurring challenges across exploration, implementation, sharing/scaling, and finalization, and second \emph{Album}, an open-source framework that implements this cycle with a minimal, executable abstraction. At Album's core is the \emph{solution}: a single Python entry-point that defines a schema-validated manifest (metadata, arguments, and environment requirements) and imperative lifecycle hooks (install/test/run) that implement the scientific routine while remaining native Python. Album executes solutions using a two-context architecture: a controller in a host environment evaluates and migrates the manifest, creates a per-solution environment, and delegates execution to that environment, where a pinned, dependency-light runner dispatches the requested lifecycle hook. Critically, the \emph{installation hook} serves as a post-environment setup mechanism, enabling reproducible execution even when a routine requires additional steps beyond dependency installation (e.g., compilation, external resources, or interactive tooling setup).
% catalogs
Album further introduces \emph{catalogs} as a decentralized, git-native distribution mechanism for solutions, augmented with an index for searchability and optional static web rendering for human-friendly discovery. This enables sharing at multiple scales, from exchanging a single solution file for rapid feedback to maintaining lab- or community-scoped catalogs, as well as publishing versioned releases (including DOI-based dissemination), while preserving provenance and governance.
% LLM interaction
Finally, Album is designed to work in conjunction with large language models (LLMs). The single-file, Python-native solution format is well-suited for LLM-assisted authoring and modification. More importantly, solutions provide \emph{executable building blocks grounded in a real inventory} that can be LLM-assisted discovered and composed (e.g., via MCP-based tool interfaces) into workflows, enabling users to shift workflow assembly from free-form command generation to tool-grounded selection and invocation of versioned solution artifacts. This enables a new collaboration mode in which domain experts can express goals in natural language while execution is anchored to validated, versioned solution artifacts. 
% examples of usage along the cycle  
We evaluate Album through four real-world deployments selected to stress distinct framework claims: the interactive visualization of beta-cell imaging data tests support for interactive, multi-tool routines that require post-environment setup \cite{muller_modular_2024}; the integration of multiple segmentation algorithms for microscopy analysis tests the composition of interchangeable backends with conflicting dependencies \cite{giese_polarity-jam_2025}; integration with workflow systems such as Snakemake for mineral quantification pipelines tests reuse of the same executable unit across interactive and workflow/HPC contexts \cite{molder_sustainable_2021, albrecht_ari3d_2025, gupta_standardized_2024, godinho_mounted_2021}; and the orchestration of a cryo-electron tomography competition tests catalog-governed dissemination and community-scale coordination \cite{harrington_open-source_2024, peck_annotating_2024}. The remainder of the paper follows the stages of the iterative cycle: search and exploration, creation and implementation, sharing and scaling, and finalization and distribution.
% glossary reference 
For terminology reference, we provide a glossary in Supplementary Table \ref{tab:glossary}.

\subsection*{Contributions}
We contribute the following. (i) We conceptualize an iterative, collaboration-driven research software cycle and derive two primitives (solution, catalog) that address recurring barriers across phases. (ii) We present Album, a two-context execution architecture with per-solution environments, schema-validated manifests, lifecycle hooks, and git-native catalogs with indexed search and optional web rendering. (iii) We perform a feasibility evaluation and demonstrate the approach across four imaging workflows, spanning interactive visualization, dependency-conflicting segmentation backends, competition pipelines, and workflow integration. (iv) Additionally, we provide illustrative examples of LLM-assisted drafting and implement a proof-of-concept MCP interface that exposes solutions as tools for tool-grounded orchestration. Overall, Album complements package managers, workflow systems, and container runtimes by making scientific routines executable, shareable artifacts.

\section*{Iterative, collaboration-driven research software cycle}\label{icdrsc}
% Leading to the iterative dev cycle
To design a framework that supports collaborative scientific software development, we first abstracted the recurring tasks that arise when domain experts and scientific developers iteratively build, test, and share research routines. Drawing on our experience and related discussions of research software ecosystems and sharing infrastructures \cite{wilkinson_ecosystem_2025, gustafsson_workflowhub_2025}, we identified four recurring phases:
% the cycle fields
\begin{enumerate}[label=\Alph*]
\item \textbf{Search \& Exploration}: find, access, and reproduce relevant prior work;
\item \textbf{Creation \& Implementation}: adapt or extend existing software for the concrete use case and validate it through testing
\item \textbf{Share \& Scale}: share routines with collaborators and users, scale execution to larger studies, and integrate routines into broader workflows and  
\item \textbf{Finalize \& Distribute}: document, package, and release routines in forms suitable for long-term reuse.
\end{enumerate}
% Describe the cycle
Figure~\ref{fig:icdssdp} summarizes this \emph{iterative, collaboration-driven research software cycle}. The phases are not strictly sequential; in practice, teams repeatedly revisit earlier phases (e.g., returning from implementation to exploration when assumptions change, or from scaling back to implementation when failures are discovered). The cycle may involve multiple developers and research groups and can also incorporate large language models (LLMs) as collaborators. It is therefore not bound to a specific time frame.
% Reasoning: looking at the cycle and identify two technical components
From this cycle, we derive the core technical requirements for a supporting system: it must (i) represent an executable scientific routine together with its environment and interfaces in a form that is easy to adapt and re-run, and (ii) provide a collaborative distribution and discovery mechanism with provenance and governance. 
% concluding what technical aspects are necessary
These requirements motivate two central primitives in Album: the \textit{solution} (the unit of execution) and the \textit{catalog} (the unit of sharing and discovery).

\subsection*{The solution file}
% intro: file
At the core of Album is the \emph{solution}. A solution is a single, user-authored Python entry-point that Album evaluates to obtain a machine-readable specification and a set of lifecycle hooks that implement a scientific routine. For the scientific developer, the solution is the primary artifact for packaging a routine for reuse and collaboration.
% intro: capture science nature
From a computational perspective, a reusable research routine should bundle three concerns. First, it should capture the scientific context: what the routine does, how it should be cited, and which parameters it expects. Second, it should capture the execution context: the dependencies and environment required to run it. Third, it should expose operational interfaces that support iterative collaboration, including installation or setup, execution, and testing or validation.
% reasoning: structure definition 
Album co-locates these concerns in a single Python entry-point. The file declares metadata and arguments, defines standardized lifecycle hooks for install/test/run, and specifies the per-solution environment. Figure~\ref{fig:album_solution_catalog_concept}A illustrates the concept; Supplementary Fig.~\ref{fig:album-single-file-view} provides a detailed view on a single Python entry-point, and Annex \ref{supplemental:solution_example} shows an example code solution.
%A minimal Python example is provided in supplement~\ref{supplemental:solution_example}.
% gap and facts
Where conventional approaches distribute this information across multiple artifacts, Album consolidates it in one executable entry-point with standardized lifecycle semantics. 
% conclusion
This is designed to reduce conceptual and operational overhead for authors and collaborators while enabling reproducible execution and post-environment setup for interactive or system-dependent routines. Consolidating metadata and the environment description in the solution file also aligns with FAIR-for-software recommendations that software metadata should include dependencies and that software dependencies be documented in a formal, accessible, machine-readable way \cite{lamprecht_towards_2020}. Because the solution remains plain Python, it is also amenable to LLM-assisted generation and modification.

\subsection*{The catalog}
\begin{figure}[ht!]
    \centering
    \includegraphics[width=0.5\textwidth]{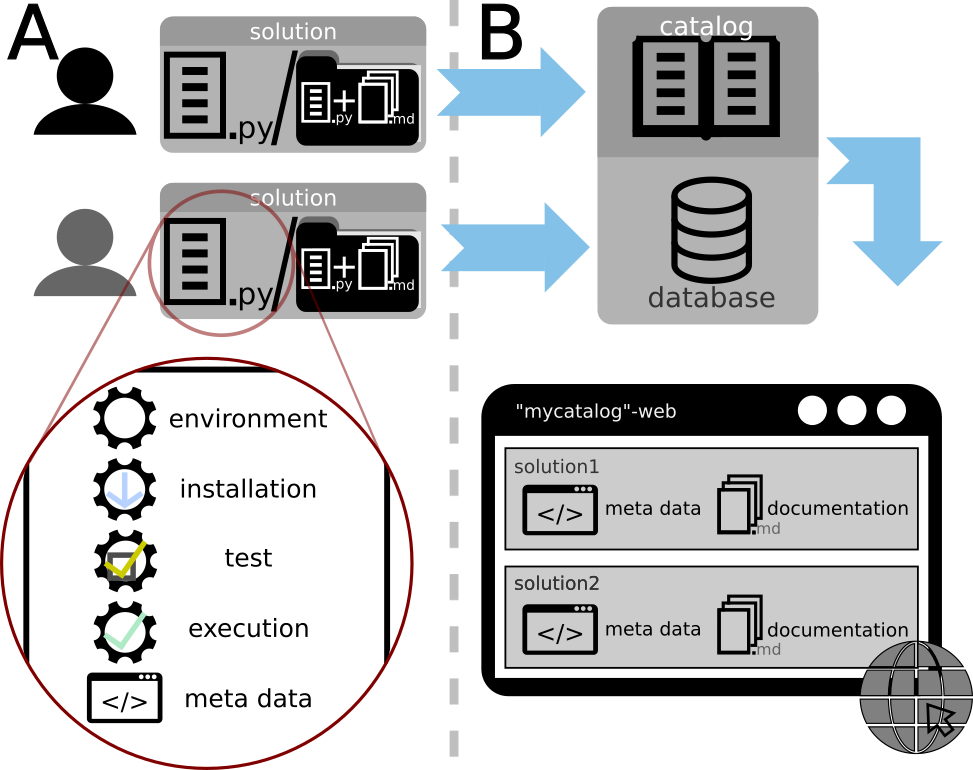}
    \caption{\textbf {Solution and Catalog concept.} {\textbf A:} Depicts the concept of a solution. A fully defined solution consists of its environment, installation, execution and test routine, as well as its metadata. {\textbf B:}  Conceptual view of a catalog. A catalog consists of a database. When hosted by any version control and collaboration platform the catalog can be rendered into a web view, where metadata and documentation are displayed in a user friendly way.}
    \label{fig:album_solution_catalog_concept}
\end{figure}

% intro: concept of a catalog
An Album \emph{catalog} is the unit for sharing and discovering solutions and responsible for the collaborative aspect of the framework. Solutions can be deployed to a catalog, where their metadata is extracted and indexed alongside the solution artifact to make routines searchable and easy to reuse.
% reasoning and gaps that are tackled in contrast to other systems
Technically, a catalog is a Git repository augmented with a structured database index. Because catalogs are plain Git repositories, they can be created and hosted anywhere, enabling decentralized sharing without requiring a centralized registry. This differs from package registries such as PyPI, where distribution is typically mediated through a dedicated registry service rather than a version-controlled repository.
% intro: catalog details
Catalogs may be hosted on standard collaboration platforms such as GitHub or GitLab, or maintained locally on disk or network storage. When hosted on platforms with CI/CD support, a catalog can additionally render a static web view from its database index, providing a browser-based interface to solutions and their documentation (Fig.~\ref{fig:album_solution_catalog_concept}B). 
% conclusion
Together, Git-native hosting, indexing, and optional web rendering provide a lightweight method for sharing solutions within a lab, a community, or publicly at a larger scale.\\

%%%%%%%%%%%%%% closing technical aspects %%%%%%%%%%%%%%
% cycling back to why both concepts exist
Our framework is organized around these two central elements, which target the core technical barriers to collaboration that recur across the four phases of the iterative cycle.
% introduce the remainding structure of the article
In the remainder of this article, we guide the reader through the iterative, collaboration-driven research software cycle as supported by Album. We illustrate how solutions and catalogs are used in each phase, drawing on diverse scientific use cases to demonstrate the approach as a proof of concept while also showcasing its feasibility and practical utility.

%%%%%%%%%%%%%%%%%%%%%%%%% Search and Explore %%%%%%%%%%%%%%%%%%%%%%%%%
\section*{Search \& Explore}
\namedsection{sec:sande}{Search \& Explore}
% intro: find
The first step for a researcher facing a computational task is to obtain an overview of available software that could address the scientific problem, either directly or as a starting point for adaptation.
% facts: find
Common discovery entry points include research software directories such as Helmholtz Research Software\footnote{\href{https://helmholtz.software/}{https://helmholtz.software/}} and Helmholtz Imaging Connect\footnote{\href{https://connect.helmholtz-imaging.de/}{https://connect.helmholtz-imaging.de/}}, container registries such as Docker Hub\footnote{\href{https://hub.docker.com/}{https://hub.docker.com/}}, and workflow repositories such as nf-core\footnote{\href{https://nf-co.re/pipelines/}{https://nf-co.re/pipelines/}}.
% gaps: find
However, these platforms are typically optimized for broad dissemination and do not directly support the rapid curation of lab-internal, frequently executed routines with consistent execution interfaces and lightweight governance.
% reasoning & conclusion: find
Album addresses this gap through catalogs. A catalog is a Git repository with an indexed view of solutions, enabling labs or communities to curate and share executable routines using familiar version-control workflows. Once a catalog is added to an Album installation, users can discover and obtain solutions via the command line or GUI. Catalogs can also be rendered as a web view to support browsing and lightweight documentation.

\begin{figure}
    \centering
    \includegraphics[width=0.5\textwidth]{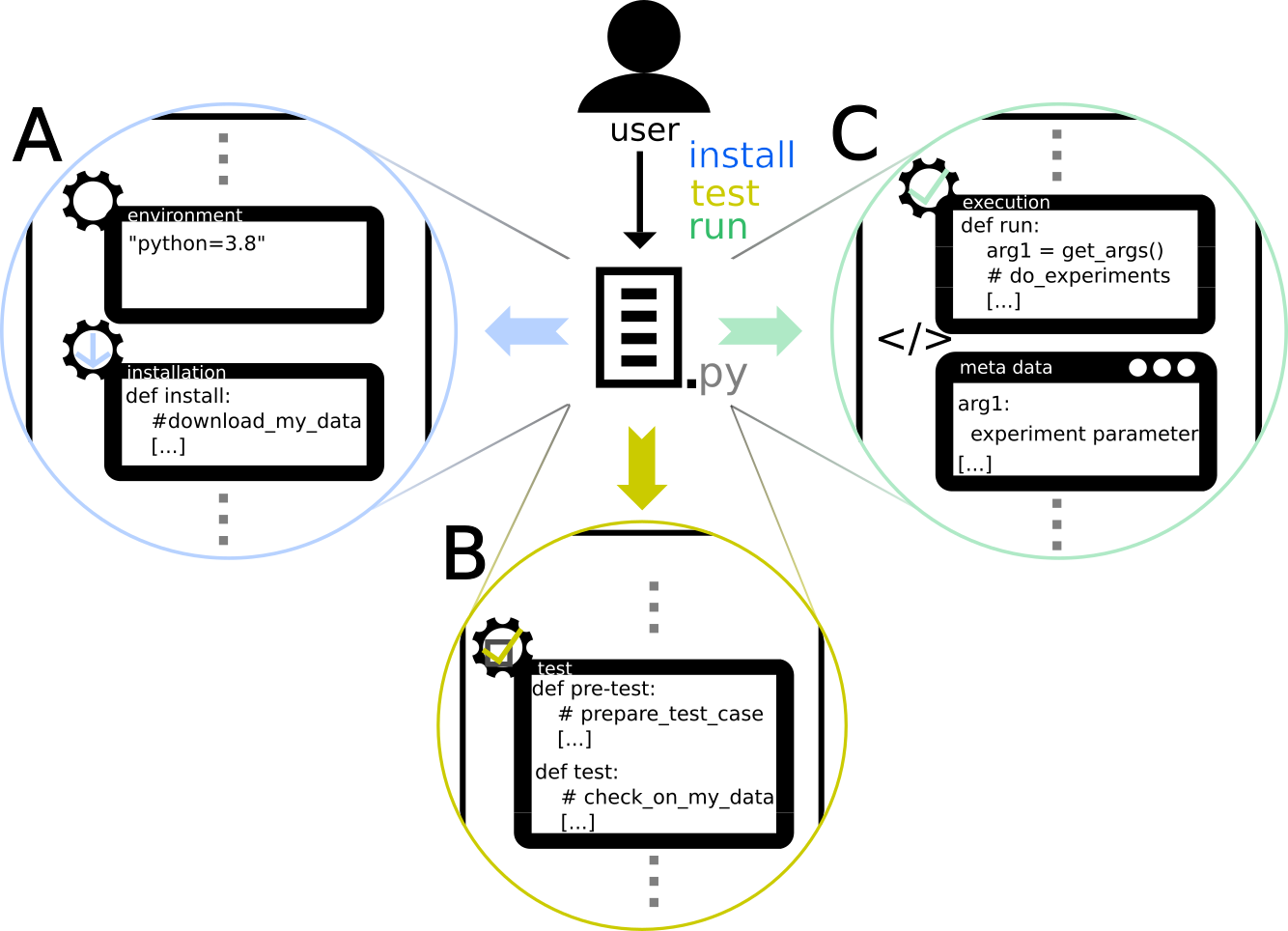}
    \caption{ \textbf{Detailed view on the definition blocks of a solution file.} Environment description, together with the installation definition (\textbf{A}). Both are used during the installation procedure of a solution. The code defined in the test block (\textbf{B}) gets executed when calling the testing functionality of a solution. Arguments defined in the metadata (\textbf{C}) can be accessed in the execution block when running a solution.}
    \label{fig:solution_detail}
\end{figure}

% intro: access
Once a candidate routine is identified, the next step is to access it and understand how to install and use it.
% facts and gaps: access
This step is often time-consuming because users must both understand how a routine is intended to be installed and executed and reproduce a compatible software environment; in practice, documentation and setup instructions are frequently incomplete or inconsistent \cite{nguyen-hoan_survey_2010}.
% reasoning: access
Album facilitates access by presenting solution metadata and a standardized install interface derived from the solution definition. During installation, Album creates an isolated per-solution environment from the solution’s environment specification, reducing accidental interference between routines and avoiding “works on my machine” drift caused by ad-hoc dependency changes. Per-solution isolation further reduces cross-contamination between routines and preserves version-specific behavior during iterative development.
% further reasoning and evidence: access
Where required, solution authors can define an installation hook that performs post-environment setup steps, enabling complex scenarios such as interactive pipelines that rely on additional tooling (e.g., Blender-based workflows) \cite{muller_modular_2024}. Fig.~\ref{fig:solution_detail}A illustrates the information involved. Installation can be initiated through either command-line or graphical interfaces.

% intro and gap: reproduce and use
After installation, users frequently encounter friction when executing research software because entry points, parameter conventions, and configuration formats vary widely across tools.
% reasoning: reproduce and use
Album makes usage explicit and executable by providing a uniform execution interface: a single command runs either the execution or testing hook defined in the solution (Fig.~\ref{fig:solution_detail}B,C). Solutions can declare arguments with types and descriptions, allowing executions to be parameterized consistently and improving repeatable re-execution for comparable use cases. In the GUI, users are prompted for the required arguments along with their descriptions (Supplementary Figs.~\ref{fig:album-gui}, \ref{fig:album-gui-install}, \ref{fig:album-gui-execution}).

% validate: example
We applied a catalog of deployed solutions in \cite{muller_modular_2024, muller_structure_2024}. In this work, a modular workflow for visualizing beta-cell data was developed, spanning volume electron microscopy processing through rendering in Blender\footnote{\href{https://www.blender.org/}{https://www.blender.org/}} (Fig.~\ref{fig:use-case}A). Wherever feasible, each modular step was exposed as a solution registered in a catalog.

% conclude: search & explore
Overall, Album is designed to reduce practical barriers in the Search \& Explore phase by making routines discoverable as executable units that include explicit environment requirements and standardized entry points. This targets common sources of overhead related to installation, permissions, and re-execution that frequently arise when adopting software from general-purpose repositories \cite{baker_1500_2016, baker_over_2015}.

% album clue card reference 
Supplementary Table \ref{tab:album-clue-card} A provides an overview of all functionalities of Album that support domain experts as well as developers of algorithms or tools during the Search and Explore phase of the iterative, collaboration-driven research software cycle.

%%%%%%%%%%%%%%%%%%%%%%%%% Creation % Implementation %%%%%%%%%%%%%%%%%%%%%%%%%
\section*{Creation \& Implementation}
\namedsection{sec:candi}{Creation \& Implementation}
% creation and implementation: intro + gap
Structuring scientific code for reuse and repeatable re-execution often receives limited attention during exploratory development, in part because it requires additional tooling and discipline beyond the immediate research goal. Yet reusing and extending existing open-source software can substantially accelerate research and innovation, especially when routines are packaged in a form that others can readily adapt \cite{tran_how_2023}.
% reasoning: solution abstraction
Album aims to reduce this overhead by providing a lightweight unit of reuse: the solution. Converting a routine into a solution makes its entry points, parameters, and environment assumptions explicit and immediately executable through Album, which in turn supports iterative refinement and enables future developers (or generative systems) to build upon the routine with fewer setup steps.
% intro: adapt
When building upon existing code, it is often necessary to adjust the execution environment, for example to add missing dependencies or to test alternative versions.
% reasoning + conclude: adapt
Album supports this iteration by allowing developers to interactively enter and inspect the per-solution environment associated with a specific routine. This supports debugging and incremental environment changes without conflating dependencies across unrelated routines.
% intro & gap: extend
In many domains, reusable functionality is embedded inside large libraries or complex applications, which can make it difficult to extract, adapt, and share a focused routine for a specific task.
% reasoning: extend
Album addresses this by enabling developers to clone an existing solution as a starting point for modification. The result is a distinct solution that can be iterated independently while retaining the standardized install/test/run structure.
To accelerate adaptation, we provide solution templates for plain Python and for common imaging ecosystems including Java, ImageJ \cite{schneider_nih_2012}, ImageJ2 \cite{rueden_imagej2_2017}, BigDataViewer \cite{pietzsch_bigdataviewer_2015}, ImgLib2 \cite{pietzsch_imglib2_2012}, SciView \cite{gunther_tales_2020}, and Napari \cite{sofroniew_napari_2025}.
% conclude: extend
Taken together, these mechanisms lower the barrier to building upon previous work and help keep execution assumptions explicit as routines evolve.
% intro: testing
Testing is important to detect regressions and to ensure that a routine continues to behave as intended as it is adapted.
% facts: testing
In conventional software projects this is typically implemented using frameworks such as unittest or pytest and its extensions \cite{liu_pytest-inline_2023}.
% gaps: testing
For many research code bases, however, the perceived overhead of adopting and maintaining full testing infrastructure can delay or prevent systematic validation.
% reasoning: testing
Album therefore offers a lightweight, solution-scoped testing convention. Solutions may define a \emph{pre-test} step to prepare inputs or fixtures and a \emph{test} step that asserts expected behavior for the routine under defined conditions.
% conclude: testing
This design is intended to lower the coordination cost of adding routine-level checks during iterative development, while remaining compatible with more comprehensive testing approaches.
% conclusion: creation and implementation
Overall, Album provides a workflow that supports iterative adaptation of focused scientific routines by isolating environments per routine, enabling cloning-based refinement, and lowering the barrier to adding routine-specific validation.\\
% point to supplement
Supplementary Table \ref{tab:album-clue-card} B outlines the functionalities of our framework that facilitate the Creation and Implementation phases of scientific work.

\section*{Leveraging LLMs with Album Solutions}
In software-heavy fields such as computer science, large language models (LLMs) appear to have become a practical tool in the research workflow at least for scientific writing \cite{liang_quantifying_2025}. In the context of Album, LLMs are useful in two complementary ways: (1) drafting or modifying Album solutions from natural-language requirements, and (2) assembling multi-step analyses by orchestrating existing solutions via the Model Context Protocol (MCP)\footnote{\href{https://modelcontextprotocol.io}{https://modelcontextprotocol.io}}.

\subsection*{Creating Album Solutions with LLMs}
% reasoning: Creating Album Solutions
Album solutions follow a constrained, machine-checkable structure that combines metadata, an explicit environment description, and lifecycle hooks for install/test/run. This structure makes solutions amenable to LLM-assisted authoring: given a clear prompt, an LLM can draft the entry-point file, including argument declarations and the corresponding hook implementations. In particular, the presence of an explicit installation hook allows LLM-generated solutions to capture post-environment setup steps when required (e.g., downloading assets or preparing external tooling).
% validate: Creating Album Solutions
In the Supplementary Materials (Annex~\ref{supplemental:prompting}), we provide example prompts that researchers can use to instruct LLMs to create or update Album solutions. Because solutions are executable units with standardized test and run entry points, the resulting artifacts can be validated through Album by executing the corresponding lifecycle hooks and inspecting logs, rather than relying on the model output alone.

\subsection*{Orchestrating Album Workflows with MCP}
% facts and gaps
Constructing complex workflows typically requires manual effort and familiarity with workflow-specific rules or domain-specific languages. 
MCP provides a standardized interface for tool use, allowing an LLM to call external systems such as Album through typed operations. We implemented an Album MCP interface that allows an LLM to:

\begin{itemize}
\item discover solutions by listing catalogs and searching within them;
\item manage catalogs by adding, updating, or removing them from a local collection;
\item install solutions and their per-solution environments;
\item execute solutions with parameters derived from natural-language requests;
\item chain multiple solutions into multi-step analysis pipelines, and
\item assist with troubleshooting by inspecting execution logs and error reports.
\end{itemize}

% validate: Orchestrating Album Workflows
In our examples (Annex~\ref{supplemental:mcp}), LLMs were able to assemble and execute solution sequences using these MCP operations. Importantly, orchestration is grounded in the catalog inventory: the LLM can only select from solutions that actually exist and can be executed by Album, which reduces reliance on free-form, non-executable suggestions.\\

% conclusion
Together, these two modes of use show how Album’s abstractions (solutions as executable units with explicit environments and catalogs as discoverable inventories) provide a structured interface for integrating LLM assistance into scientific software development and workflow composition.

%%%%%%%%%%%%%%%%%%%%%%%%% Share and Scale %%%%%%%%%%%%%%%%%%%%%%%%%
\section*{Share \& Scale}
\namedsection{sec:sands}{Share \& Scale}
% intro: sharing
Collaborative work can improve research outcomes \cite{liao_how_2011, vaudano_research_2020}, and collaborations that bridge domain science and software development often yield reusable community contributions \cite{muller_modular_2024, giese_polarity-jam_2025}. In practice, however, collaboration requires sharing routines at different stages of maturity and with different audiences: from a single collaborator during prototyping to a broader community and, eventually, to archival publication.

% gap: sharing modes
Many existing distribution mechanisms are optimized for one of these stages (e.g., ad-hoc file sharing, developer-centric repositories, package registries, or archival snapshots), but do not natively support a smooth progression between them while keeping a routine executable and well-scoped throughout.

\begin{figure}
    \centering
    \includegraphics[width=0.5\textwidth]{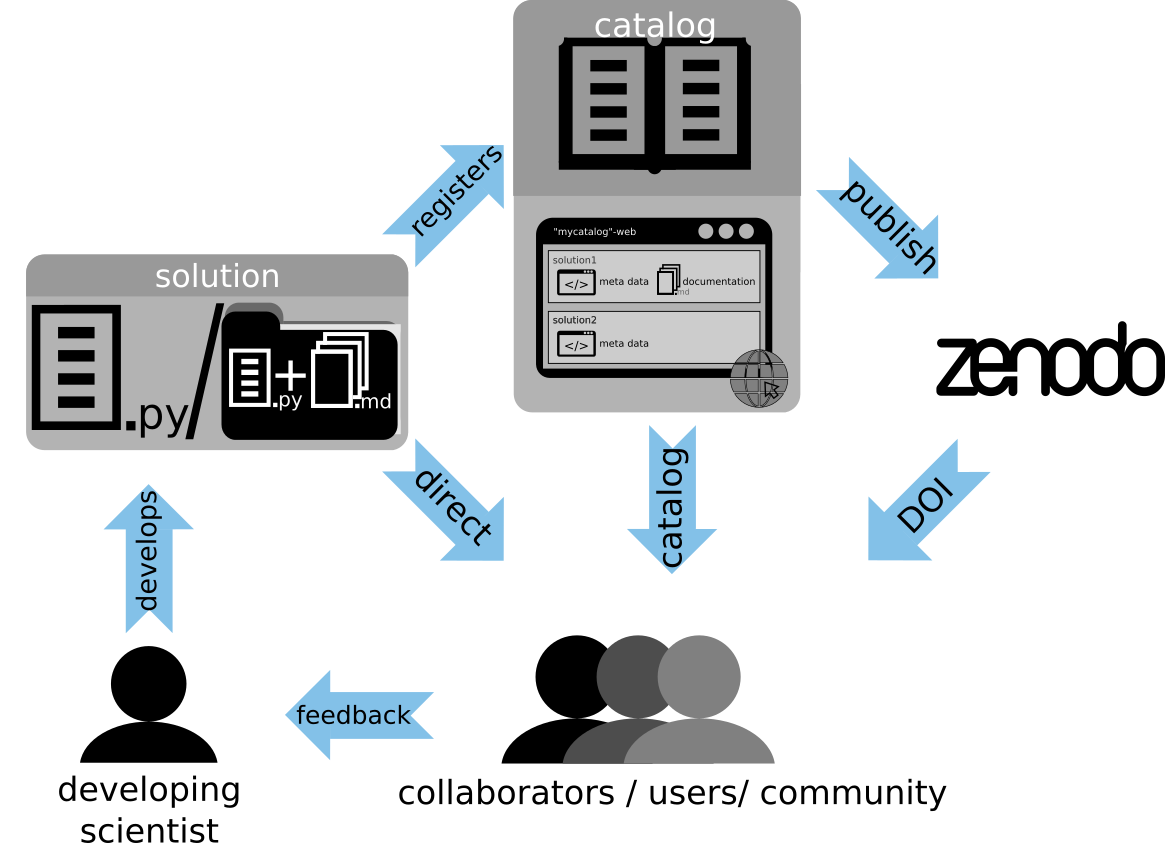}
    \caption{ \textbf{Shareability concept.} There are several ways to share a solution. One option is to directly send the solution or its folder to a collaborator, which is useful for prototyping or gathering quick feedback. Another approach is deploying the solution to a catalog, allowing users to access it. Lastly, the solution can be published to a catalog that integrates with Zenodo for broader dissemination.}
    \label{fig:sharibility}
\end{figure}

% reasoning: sharing (three modes)
Album supports staged sharing through three complementary modes.
First, a solution file (or solution directory) can be shared directly and executed by Album without a catalog, which is designed to enable fast iteration with a small set of collaborators.
Second, solutions can be deployed to catalogs hosted on Git-based platforms (e.g., GitHub or GitLab), leveraging familiar contribution workflows (issues, merge/pull requests) and platform-level access control for lab- or community-scoped sharing. Platforms that have been shown to enhance collaborative and reproducible research \cite{braga_not_2023}.
Third, catalogs can integrate with Zenodo to archive released versions and associate them with DOIs, enabling citation and long-term referencing of specific solution snapshots (Fig.~\ref{fig:sharibility}). Album can resolve Zenodo-backed DOIs to retrieve the archived artifact and treat it as an executable solution.

% further reasoning :sharing - execution semantics
Because solutions expose standardized install/test/run entry points and declared arguments, collaborators can communicate execution instructions consistently (e.g., “install and run solution X with these parameters”) 
through a standardized interface compared to ad-hoc scripts.

% validate: community catalog
As an example of community sharing, the Helmholtz Imaging Catalog\footnote{\href{https://album-app.gitlab.io/catalogs/helmholtz-imaging/}{https://album-app.gitlab.io/catalogs/helmholtz-imaging/}} curates solutions developed through collaborative community efforts and is integrated into Helmholtz Imaging Connect\footnote{\href{https://connect.helmholtz-imaging.de/}{https://connect.helmholtz-imaging.de/}} to increase visibility and browseability.

% validate: cellcanvas
Moreover, the cellcanvas catalog\footnote{\href{https://album.cellcanvas.org/catalog}{https://album.cellcanvas.org/catalog}} was used as part of the tooling for a Kaggle competition on CryoET object identification\footnote{\href{https://www.kaggle.com/competitions/czii-cryo-et-object-identification/overview}{https://www.kaggle.com/competitions/czii-cryo-et-object-identification/overview}} based on the realistic phantom dataset described by Peck et al. \cite{peck_realistic_2025}.
\newcounter{fnx}% 
\setcounter{fnx}{\value{footnote}}
In this setting, Album solutions supported multiple steps ranging from ground-truth generation to interactive tooling (e.g., integration with napari \cite{sofroniew_napari_2025} and Copick\footnote{\href{https://copick.github.io/}{https://copick.github.io/}}) and competition simulation/optimization (Fig.~\ref{fig:use-case}D). Participants reported in discussion threads the stability of the tooling and workflow\footnote{\href{https://www.kaggle.com/competitions/czii-cryo-et-object-identification/discussion/561433}{https://www.kaggle.com/competitions/czii-cryo-et-object-identification/discussion/561433}}.

% conclude: sharing
Together, these examples illustrate how Album supports sharing focused, executable routines from early collaboration through community dissemination and archival publication.

%%% scaling - locally
% intro: scaling
After a routine has been shared and reproduced, a common next step is to scale it from a minimal test case to larger datasets or repeated runs.
% gap: scaling - batch run locally
For many routines, scaling begins with simple batch execution (e.g., iterating over files or parameter sweeps), which otherwise requires custom scripting and manual bookkeeping.
% reasoning & evidence: scaling - batch run locally
Album provides a batch-execution plugin that supports running solutions over collections of inputs. Early versions of this approach were used in \cite{finotto_singlecell_2023} to enable batch processing for macrophage detection in 3D confocal images using Fiji.
% conclude: scaling - batch run locally
This reduces the effort required to scale a solution from single-case execution to repeated runs.

%%% scaling - with HPC
% intro: scaling - with HPC
Scaling can also involve running routines in high-performance computing (HPC) environments and composing multiple steps into larger pipelines.
% facts: scaling - with HPC
Workflow managers are commonly used for this purpose \cite{wratten_reproducible_2021}, but end-to-end robustness remains challenging in practice \cite{marchment_bioflow-insight_2024}.
% reasoning & evidence: scaling - with HPC
We demonstrate integration of Album solutions with Snakemake \cite{molder_sustainable_2021}, showing that the same solution can be reused as a workflow step on HPC while also remaining executable interactively in a local environment \cite{albrecht_ari3d_2025} (Fig.~\ref{fig:use-case}C).
% conclude
This illustrates reuse of the same executable unit across interactive and workflow-driven contexts.

%%% integrate
% intro: integrate definition
Beyond workflow composition, developers often need to integrate routines programmatically into a larger application while isolating incompatible dependencies.
% gap: integrate - version conflicts
In practice, different algorithms may require conflicting dependency versions, making it difficult to combine them in a single environment.
% reasoning: integrate - multi-environment execution
In \cite{giese_polarity-jam_2025}, Album was used to incorporate segmentation routines based on Mesmer/DeepCell \cite{greenwald_whole-cell_2022} and MicroSAM \cite{archit_segment_2025}, which have differing dependency constraints (Fig.~\ref{fig:use-case}B). Album enabled these routines to be executed in separate per-solution environments and invoked from a shared Python context, reducing the risk of dependency conflicts. 
We treated solutions as interchangeable backends behind a stable semantic interface. Results were exchanged through a backward-compatible interface between the calling code and the solution outputs. This allowed a host application to compose multiple backends without first merging their dependencies into a single shared environment. Instead, each backend remained a separately installable and executable solution, preserving dependency isolation while supporting interactive integration. For an illustrative comparison of coordination overhead across approaches, see Annex \ref{supplemental:work_comparison} and Supplementary Table 4.
% conclude: integrate - update problem + catalogs
When solutions are used as modular components, they must also be maintainable and updatable. Once a larger tool composes multiple routines that execute in separate, potentially incompatible environments, maintenance becomes a vital problem: individual routines must be updatable without forcing a rebuild of a monolithic, shared environment and without destabilizing other steps.
% reasoning: integrate - update problem 
Album addresses this by treating each routine as a versioned, addressable solution (e.g., \texttt{<catalog:group:name:version>}) distributed through catalogs. Updating a single step is expressed as releasing a new solution version (with its own environment specification and hooks) and deploying it to the catalog; downstream applications or workflows can pin the previous version for stability or selectively upgrade by switching to the newer solution identifier. Because execution occurs in per-solution environments, dependency changes remain localized to the updated routine.
% evidence
We applied this pattern in \cite{albrecht_ari3d_2025}, where workflow automation relied on catalog updates to evolve individual steps over time while retaining executable, versioned references for each routine.

% conclude: share \& scale
Overall, Album combines per-solution environment isolation with standardized lifecycle semantics and catalog-based versioning, enabling routines to be shared, scaled, and integrated as modular building blocks in both interactive and workflow-driven contexts. This modular approach shifts some burden to stable interchange formats and I/O boundaries between steps.

Supplementary Table \ref{tab:album-clue-card} C outlines the specific functionalities of Album that can be employed to address the identified issues.

%%%%%%%%%%%%%%%%%%%%%%%%% Finalize and Distribute %%%%%%%%%%%%%%%%%%%%%%%%%
\section*{Finalize \& Distribute}
\namedsection{sec:fandd}{Finalize \& Distribute}
% intro
At the end of a full pass through the iterative, collaboration-driven research software cycle, routines often transition from exploratory prototypes to artifacts that others can re-run, inspect, and cite. In this finalization phase, the goal is to stabilize documentation and distribution in ways that improve replicability and, ideally, reuse over time. In later iterations of the cycle, this phase also includes maintenance-oriented updates.

% gap: documentation
Retaining niche operational knowledge (e.g., instrument-specific quirks or analysis conventions) is difficult in practice, and such knowledge is frequently lost when researchers leave.

% reasoning: documentation
Album supports attaching richer documentation by allowing solutions to be authored in a directory-based form that includes additional artifacts (e.g., Markdown documentation alongside the Python entry-point). When solutions are distributed via a catalog and the catalog is configured to render a static web view (e.g., via CI/CD on common hosting platforms), this documentation becomes browsable alongside solution metadata.

% conclude: documentation
This provides a lightweight mechanism to preserve and disseminate institutional knowledge together with an executable routine.

% intro: replicability / packaging
Documentation alone is insufficient if a routine cannot be re-executed reliably. For longer-term replicability, additional packaging and environment-freezing mechanisms are commonly used (e.g., version pinning, lock files, or containers), but adopting them can require substantial tooling knowledge, especially for interactive routines or those that depend on specialized hardware.

\begin{figure}
    \centering
    \includegraphics[width=0.5\textwidth]{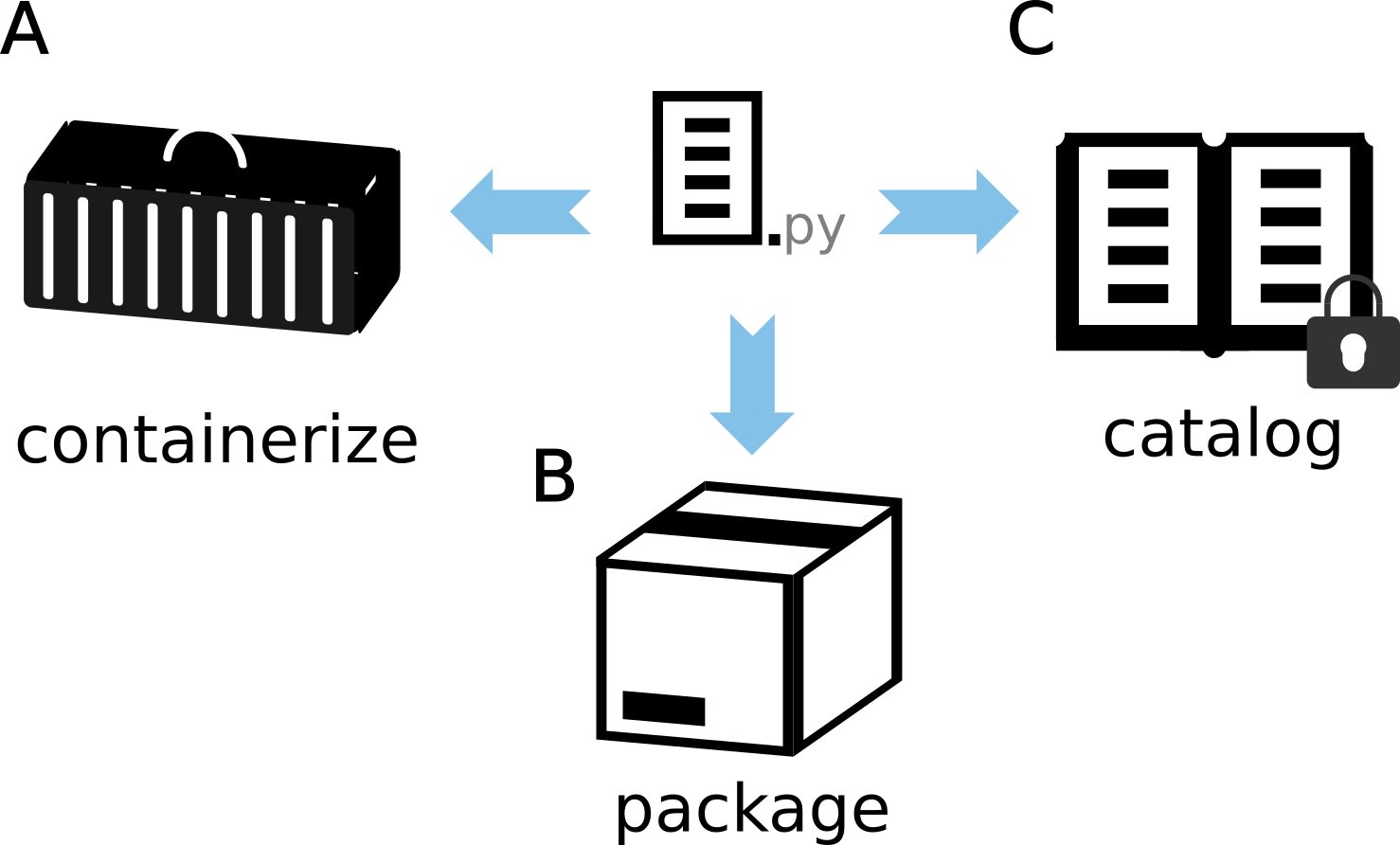}
    \caption{ \textbf{Finalization concept.} The three ways to finalize a solution. First option is to use the album-docker plugin to transfer your solution into a docker container (\textbf{A}). Second, use the album-package plugin to package your solution into an executable (\textbf{B}). And third, generate lock files via conda-lock and deploy to a public catalog \textbf{C}.}
    \label{fig:finalize}
\end{figure}

% reasoning: finalize options
Album provides multiple paths to finalize a solution, depending on the intended audience and longevity requirements.
First, solutions can be containerized via the \texttt{album-docker} plugin, producing a container image that captures the runtime environment and reduces dependency drift, as demonstrated recently \cite{albrecht_ari3d_2025}.
Second, solutions can be packaged into a distributable executable via the \texttt{album-package} plugin, which can simplify sharing with less technical collaborators (typically in a platform-specific form).
Third, Album supports environment locking via \texttt{conda-lock}, generating platform-resolved lock files that pin dependency versions and improve repeatability across installations.
Finally, solutions can be deployed through DOI-enabled catalogs that interface with Zenodo, providing citable, immutable snapshots for provenance and long-term referencing (Fig.~\ref{fig:finalize}); this improves traceability and retrieval, and complements (rather than replaces) technical mechanisms such as locks or containers.

% conclude
Together, these options provide a practical ladder for moving from prototype to distributable artifact, allowing authors to choose the level of stabilization appropriate for their collaborators and reuse scenario. We refer to Supplementary Table~\ref{tab:album-clue-card}D for the corresponding Album functionalities.

\section*{Discussion}

% summary
Album is designed around two abstractions, \emph{solutions} and \emph{catalogs}, to reflect how research software is developed in practice: iteratively, collaboratively, and under heterogeneous technical constraints. Rather than introducing a new language or replacing existing infrastructure, Album provides a lightweight executable packaging unit (the solution) and a decentralized distribution mechanism (the catalog). Together, these abstractions support searching, re-executing, adapting, sharing, scaling, and stabilizing scientific routines through its design. Below, we discuss implications of this design, the trade-offs it introduces, and concrete directions for future work.

\subsection*{Reproducibility \& replicability}
A central expectation in scientific computing is that analyses can be re-executed and verified. Research software, as defined by the FAIR4RS working group, includes ``source code files, algorithms, scripts, computational workflows, and executables that were created during the research process or for a research purpose'' \cite{barker_introducing_2022}. Following \cite{cohen-boulakia_scientific_2017}, \emph{replicability} can be understood as technical robustness: re-running the same routine under similar computational conditions yields consistent results. \emph{Reproducibility}, in contrast, concerns the stability of scientific conclusions under different tools, data, or methodological choices \cite{cohen-boulakia_scientific_2017} (noting that terminology varies across fields).

% argument why album is connected and how
Album primarily targets the technical side of \emph{replicable re-execution}: solutions bundle a machine-readable manifest, environment descriptions, standardized execution semantics, and (optionally) tests. Export mechanisms (e.g., environment locking and containerization) further support re-running routines beyond the original development context. These mechanisms can facilitate downstream efforts toward scientific reproducibility by making analyses easier to inspect, re-execute, and compare; however, they do not by themselves guarantee reproducible scientific conclusions. We evaluate whether solutions that rely on older album\_solution\_api versions remain executable (see Annex \ref{supplemental:re_executability}). With our pinned runner approach and version-appropriate dependency constraints, all tested solutions remained executable.

\begin{figure*}[ht!]
    \centering
    \includegraphics[width=1\textwidth]{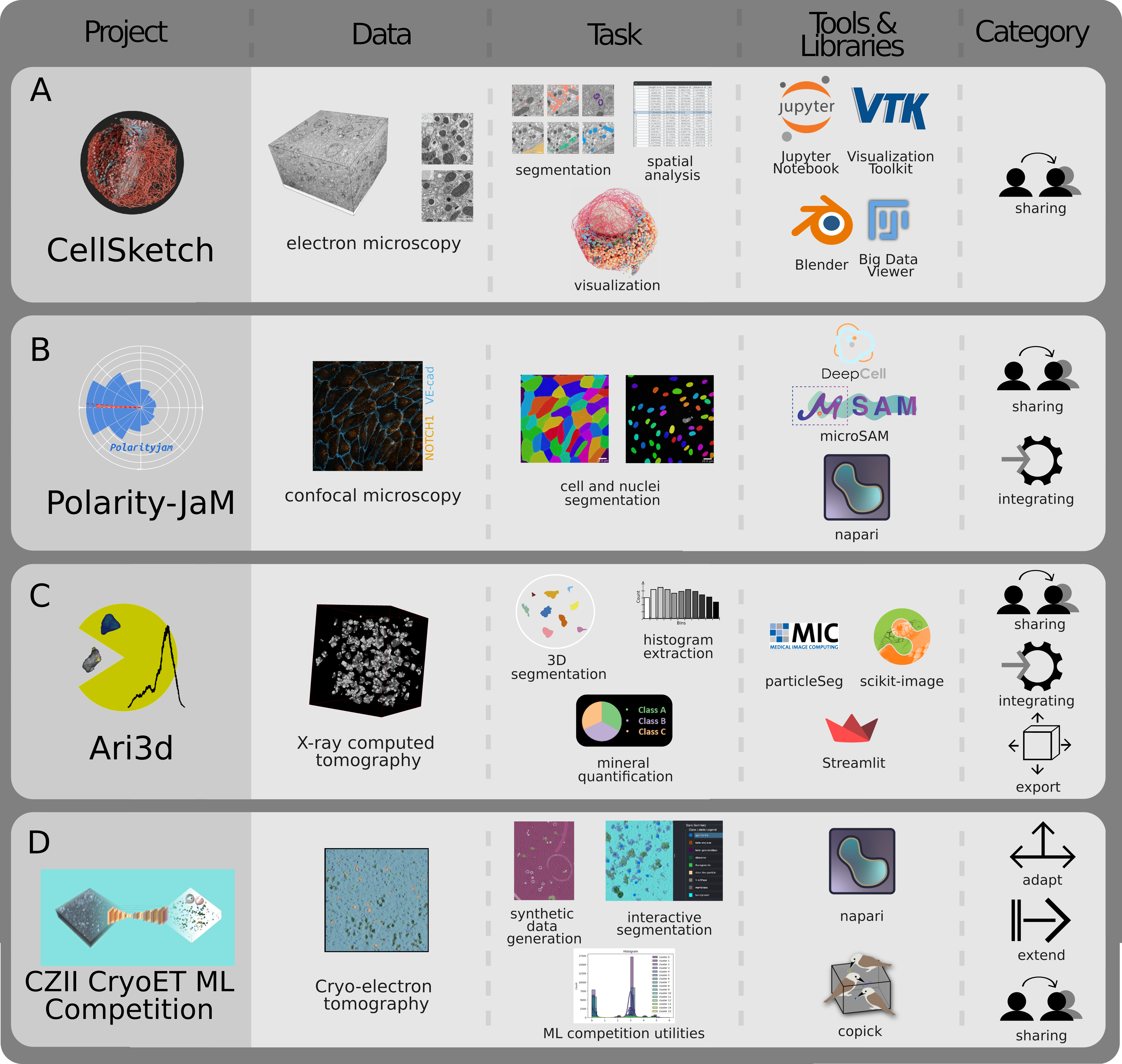}
    \caption{\textbf{Deployment-based evaluation of Album across four imaging scenarios.} We assess feasibility through real-world deployments selected to exercise the framework’s core claims across the iterative research software cycle (Fig. 1): (\textbf{A}) interactive, multi-tool visualization pipelines requiring post-environment setup. Specifically, Album was utilized to implement and disseminate a protocol for converting raw electron microscopy data into a three-dimensional visualization, thereby covering steps such as segmentation of cellular compartments, spatial analysis, and three-dimensional rendering; (\textbf{B}) programmatic composition of interchangeable segmentation backends with conflicting dependencies, e.g. segmentation algorithms for nuclei and cell detection within the API design of Polarity-JaM, a tool for studying cell polarity, junctions, and morphology. (\textbf{C}) reuse of the same executable unit in both interactive and workflow/HPC contexts and to distribute solutions required for a workflow that addresses the task of mineral quantification in three-dimensional X-ray computed tomography images. Additionally the Album-docker plugin was used to containerize the entire workflow solution; and (\textbf{D}) catalog-governed dissemination in a competition setting where Album was extensively used to facilitate a CryoET Machine Learning Competition\protect\footnotemark[\value{fnx}]. Consequently, Album solutions supported the creation of ground truth data, optimized models, facilitated the interactive use of diverse machine learning tools, and supported the simulation and enhancement of the competition.   
    } 
    \label{fig:use-case}
\end{figure*}

\subsection*{Openness and FAIR principles}
For research software to benefit the community, it must be discoverable, accessible, and reusable within the social and technical practices of a field. FAIR principles \cite{wilkinson_fair_2016} and their adaptations to research software motivate practices that improve findability, access, and reuse, but implementing FAIR-aligned workflows in day-to-day scientific development imposes overhead. Album operationalizes FAIR aspects by co-locating metadata, environment descriptions, and executable entry points, and by making catalogs indexable and web-browsable. At the same time, Album deliberately does not enforce domain-specific interoperability rules. As a result, \emph{interoperability}, especially with respect to data exchange semantics, remains a community and domain responsibility (see Supplementary Table~\ref{tab:fair4rsalbum}).

% openness of album and catalogs
Open software can significantly benefit researchers by increasing visibility, enabling collaboration, and improving impact \cite{mckiernan_how_2016}. Album is released under a permissive license (MIT) and built on open tooling to maximize long-term accessibility and reuse, aligning with the broader goals of Open Data, Open Code, and Open Infrastructure \cite{wilkinson_applying_2025}. In practice, public catalogs, community review, and clear provenance practices are essential to translate technical openness into scientific reuse.

\subsection*{Scientific infrastructure and positioning}
Album does not aim to replace package managers, workflow DSLs, or container runtimes (Supplementary Table~\ref{tab:tool_comparison}); instead, it composes with them by making an executable routine a shareable artifact with standardized lifecycle semantics. Package managers provide robust dependency distribution, but they typically leave execution entry points, post-install setup, and validation conventions to individual projects, which can add coordination overhead during collaborative prototyping, especially when collaborators have heterogeneous technical expertise. Plugin ecosystems (e.g., napari/ImageJ/Fiji) support interactive integration, but extensions usually execute within a shared environment and can become brittle when methods require conflicting dependency stacks. Workflow engines such as Snakemake, Nextflow, and Galaxy provide strong reproducible composition once steps are formalized, but are often less aligned with ad-hoc interactive iteration. Containers offer strong replication for mature artifacts, yet can be comparatively heavy to build and iterate on for interactive or rapidly evolving routines.

Album therefore defines a different unit of reuse: solutions, i.e., Python-native entry points that co-locate machine-readable metadata, an explicit environment specification (including non-Python dependencies via conda-compatible tooling), and lifecycle hooks for install/test/run. Optional post-environment installation hooks capture additional setup steps, such as downloading assets or compiling components when required. This unit enables patterns in a single shared Python environment, such as exposing multiple segmentation backends with clashing dependencies behind a stable semantic interface, while retaining a uniform execution contract and catalog-based versioning/governance. This design prioritizes practical coverage of heterogeneous scientific stacks and staged collaboration over enforcing a single packaging standard.
%Album is not a package manager, a workflow DSL, or a container platform (see Supplementary Table~\ref{tab:tool_comparison} for a comparison to UV, workflow managers, and containerization tools). Instead, Album composes with these systems by turning an executable routine into a first-class artifact with standardized lifecycle semantics. Package managers provide stable dependency resolution and distribution channels, but the cost of packaging and publishing can be prohibitive during early prototyping—especially across heterogeneous collaborator skill levels. Album addresses this gap by supporting early sharing as a single executable entry point, while enabling later stabilization through catalogs, versioning, and export mechanisms.

% album along the development cycle
%Because solutions are native Python entry points, Album remains compatible with established scientific libraries, while supporting environments that include non-Python dependencies via conda-compatible tooling. Optional post-environment installation hooks allow authors to perform additional setup steps (e.g., downloading assets, compiling components) when required. This design prioritizes practical coverage of scientific stacks over enforcing a single packaging standard.

\subsection*{Encapsulation and isolation}
Containers remain a key mechanism for long-term portability and reproducibility \cite{moreau_containers_2023}. However, building and maintaining containers, particularly for GPU workloads or interactive applications, can require specialized expertise and may reduce portability \cite{zhou_containerization_2023}. Album’s primary contribution is earlier in the lifecycle: it promotes \emph{solution-scoped isolation} via per-solution environments, which reduces cross-project dependency conflicts and supports iterative prototyping among collaborators. For longer-term archiving and deployment, Album provides pathways to containerization and DOI-based snapshots, which complement but do not replace broader reproducibility practices (e.g., archival of input data, model weights, and hardware-specific constraints).

\subsection*{Workflows and composability across environments}
Workflow systems (e.g., Snakemake \cite{molder_sustainable_2021}, Galaxy \cite{giardine_galaxy_2005}, Nextflow \cite{di_tommaso_nextflow_2017}) encode multi-step analyses in a reproducible manner, often relying on containers. They can also introduce entry barriers for non-expert users and may be brittle when individual steps are not independently reusable or testable. Album complements workflow engines by making \emph{individual steps} easier to package, execute, and validate as solutions, and by enabling those steps to run in separate, conflict-free environments.

A key implication of this design is a shift in where complexity resides: \emph{multi-environment composition} reduces dependency conflicts but increases reliance on explicit interface contracts between steps. File formats, schemas, metadata, and provenance mechanisms are well-established; however, when routines are executed in separate environments (and potentially orchestrated automatically), these boundary definitions become operationally central rather than incidental. Album intentionally does not enforce a universal output schema to accommodate interactive applications whose outputs may be user-driven (e.g., napari-based annotation). Instead, Album provides typed argument declarations and solution-scoped cache/data paths to support file-based interchange when solutions are composed programmatically. In our deployments, composition relied on lightweight, domain-specific boundary contracts (e.g., a fixed output directory within the solution cache and a small set of agreed file artifacts such as a segmentation mask stored as \texttt{.npy}). This illustrates that Album enables explicit boundary contracts without imposing a single global standard.

\subsection*{Lifecycle hooks as a minimal operational contract}
Album standardizes lifecycle hooks (installation/setup, execution, and testing) to provide a minimal operational interface for reusable research routines. We argue that this is a pragmatic lowest common denominator: it is expressive enough to capture common scientific needs (including post-environment setup and validation) without requiring a domain-specific language. Adding more hook types risks recreating workflow-specific semantics within the framework. A practical direction forward is to keep the hook set minimal while enabling extensibility through plugins and domain templates, and to strengthen conventions around arguments, logging, and I/O contracts rather than expanding the core lifecycle vocabulary.

\subsection*{Implications for LLM-assisted development and execution}
Album’s solution abstraction aligns with LLM-assisted workflows in two ways. First, a single Python entry point with a machine-readable manifest is amenable to generation and modification by LLMs, while remaining directly executable and testable by collaborators. Second, MCP-based tool interfaces allow LLMs to discover, install, and execute \emph{cataloged} routines rather than inventing ad hoc commands. This provides a practical grounding mechanism: the model selects from an explicit inventory of executable solution artifacts and invokes typed tool operations rather than emitting free-form shell commands, which is an important aspect given that hallucinations remain a known failure mode in LLM outputs \cite{li_halueval_2023}. This does not eliminate error: incorrect parameterization, unsuitable tool choice, or silent scientific failure modes remain possible.  Album exposes tool-grounded building blocks via MCP; workflows suggested by an LLM can be constrained to those blocks. This design is intended to mitigate a class of free-form hallucinations (e.g., nonexistent tools/flags), but we do not claim a measured reduction in hallucination rate in this work.

These observations suggest concrete next steps: semantic search over catalogs using embeddings of manifests and documentation; automated plan-then-validate execution that routinely runs solution tests or lightweight checks; and richer metadata that helps an LLM select among near-duplicate routines (e.g., expected input modalities, output schema, and known limitations).

\subsection*{Toward cross-domain prototyping and catalog federation}
Catalogs enable decentralized sharing and local curation. "Catalogs of catalogs" could improve discovery at institutional or community scale. However, cross-domain reuse requires more than discovery: it requires shared conventions for what constitutes compatible input/output, how results are represented, and how assumptions are communicated. A promising direction is to treat interoperability as an explicit layer: domain communities can define canonical interchange formats and provide adapter solutions that translate between representations. Album can then serve as the execution and distribution substrate for these modular translation steps, enabling cross-domain prototyping while keeping scientific assumptions explicit.

\subsection*{Security and Trust}
Album solutions are executable code: installing or running a solution may execute arbitrary instructions with the user’s permissions. Album’s per-solution environments isolate dependencies but do not provide a security sandbox. Trust, therefore, rests primarily on catalog governance and artifact provenance: users should prefer curated catalogs (e.g., maintained by labs, institutes, or working groups) and use review-gated contribution workflows for community catalogs. 
Because Album evaluates the solution entry point to extract the manifest, solution authors should avoid side effects at import time. For automated execution (including LLM-driven tool use via MCP), additional guardrails are important (confirmation prompts, allowlists of trusted catalogs, and machine-readable trust signals such as CI/test status and provenance metadata). In this work, we focus on feasibility and collaborative workflows; strengthening trust and safety mechanisms is an explicit direction for future work. Please also note that Album does not make LLM execution safe; it constrains tool choice to known artifacts.

\subsection*{Limitations and outlook}
Album provides a standardized interface for collaborative prototyping and routine reuse, but it inherits ecosystem constraints: environment resolution depends on continued package availability; platform-specific issues remain; and isolated execution increases I/O and storage demands when composing multi-step pipelines. While Album provides technical backward compatibility for catalog metadata via schema migration, larger deployments still require \emph{content governance}: review and acceptance policies for new solutions, explicit deprecation and replacement mechanisms, and machine-readable trust signals (e.g., tests/CI status, provenance, pinned runner versions) that become especially important for automated execution, including LLM-driven workflow assembly.

\subsubsection*{Integration with data management systems}
Catalogs expose structured, queryable metadata link executable routines to documentation and environment descriptions. This creates a natural integration point for institutional data management platforms. For example, in bioimaging, systems such as OMERO \cite{allan_omero_2012} provide dataset-centric organization, access control, and provenance. A promising direction is to surface Album solutions \emph{alongside} datasets in such platforms, enabling users to discover "routines applicable to this data type" and execute them with standardized semantics. Realizing this requires explicit interface contracts: well-defined input/output representations, metadata conventions, and provenance capture across solution boundaries.

\subsubsection*{Remote execution backends for specialized compute}
Many routines increasingly rely on specialized hardware (e.g., GPUs) that is not available on a collaborator’s workstation. Because Album separates control (discovery/installation) from execution (per-solution environments and hooks), it provides a conceptual basis for dispatching solutions to remote execution backends. For instance, infrastructures such as the Helmholtz Model Zoo\footnote{\href{https://hmz-hub.desy.de/}{https://hmz-hub.desy.de/}} could host GPU-capable execution services; solutions could act as executable, versioned adapters that package model invocation and post-processing consistently. Implementing this requires secure credential handling, data locality strategies, and explicit provenance reporting, but aligns with the broader goal of making advanced methods accessible without forcing users to reproduce complex environments locally.

\subsubsection*{Toward research objects and credit-aware modularity}
Modularizing complex workflows is a growing trend in workflow systems \cite{soiland-reyes_making_2024}. In Album’s ecosystem, modular solutions can support credit attribution and provenance (e.g., credit printouts, DOI snapshotting) and can align with emerging packaging approaches for computational artifacts and metadata such as RO-Crates and WorkflowHub \cite{gustafsson_workflowhub_2025}. This direction reinforces the view of solutions as reusable, citable building blocks, while again emphasizing that practical interoperability depends on shared I/O conventions.\\

Overall, Album introduces a practical layer between libraries, package managers, containers, and workflow engines. It can act as an executable \emph{digital library} of scientific routines: solutions are cataloged, searchable, and runnable with standardized semantics. The same abstraction supports early-stage collaboration (sharing an executable routine as a file) and later-stage stabilization (catalog distribution, DOI snapshotting, and containerization) without forcing a DSL or a container-first workflow. We believe the solution abstraction is compatible with a future in which scientific computation is assembled from tested, executable blocks (potentially planned by LLMs) provided that communities invest in stable interface contracts and provenance-aware composition. Consequently, the framework and the underlying concepts could be implemented more broadly, extending beyond the domain of image processing.

%\todo[inline]{easy communication of a snapshot of an execution of scientific work? e.g. "i ran this tool with this parameters - here take my solution"?}
%\todo[inline]{selling point "science package manager"?}
%\todo[inline]{argument in discussion: adapt functions to domain specific needs - see polarityjam "prepare, segment" modularization}
%\todo[inline]{guideleine for an assembled workflow into container?}
% outlook
% different base container
%- user developer interaction ?
% model context protocol? decorator to exposing that function to claude

%\todo[inline]{workflowhub? \cite{gustafsson_workflowhub_2025} also RO-Crates are intersting. this parper \cite{marchment_bioflow-insight_2024} asess ability of ro-crates to describe nextflow workflows.}

%\todo[inline]{ \cite{freire_reproducibility_2016}  PRIMAD Model is intersting }
%\todo[inline]{\cite{marchment_bioflow-insight_2024} focuses on re-se in workflows}

\section*{Methods}
\begin{figure*}[ht!]
    \centering
    \includegraphics[width=1\textwidth]{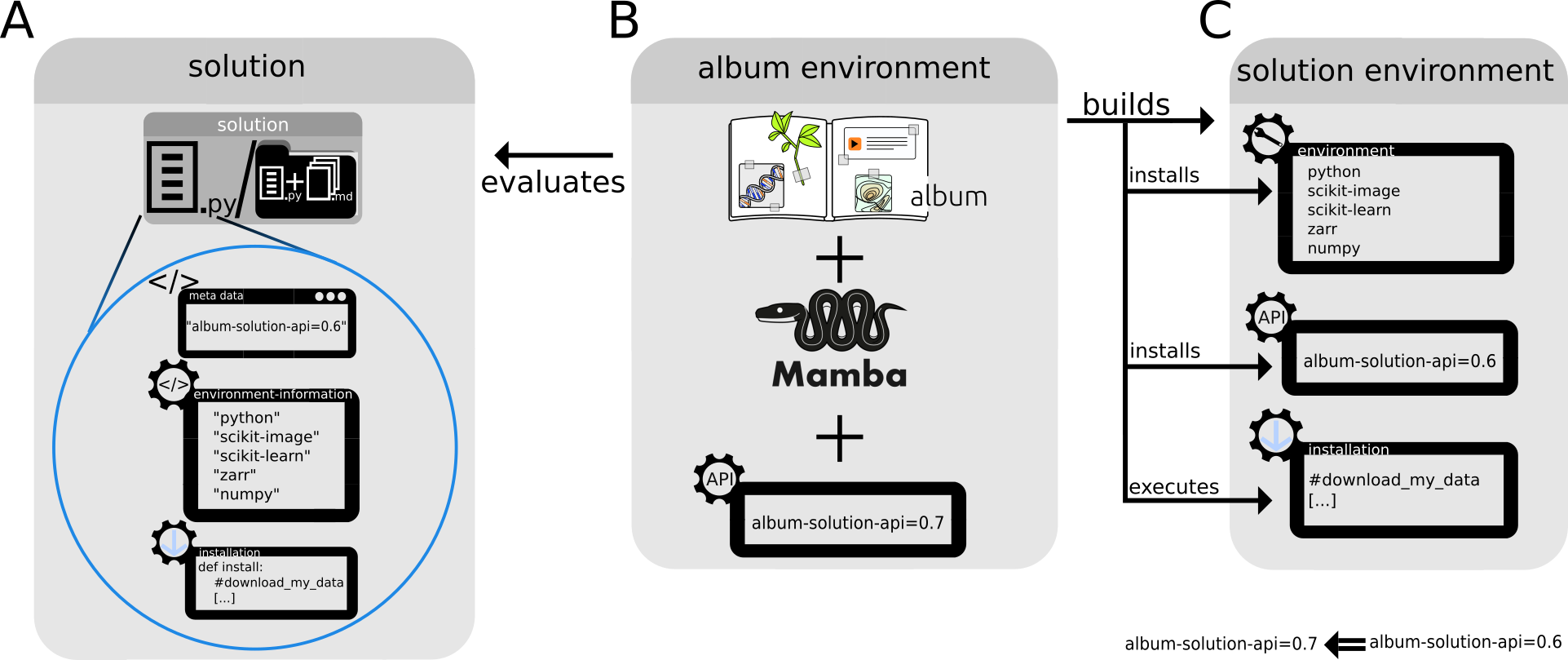}
\caption{\textbf{Album runtime principle during installation.}
(\textbf{A}) A solution entry-point defines an environment specification, an installation hook (as a lifecycle-hook definition), and a pinned runner (\textit{album-solution-api}) version.
(\textbf{B}) In the host (controller) environment, Album evaluates the solution entry-point to obtain and validate its manifest and to determine the required per-solution environment.
Album then (\textbf{C}) creates the per-solution environment using micromamba, installs the pinned \textit{album-solution-api} version into that environment, and executes the installation hook within the per-solution environment.
The \textit{album-solution-api} acts as an in-environment runner that evaluates the solution entry-point in-context and dispatches the requested lifecycle hook. The figure also illustrates that the host controller may be newer than the pinned runner version, enabling backward-compatible execution semantics across versions.}
    \label{fig:runner-principle}
\end{figure*}

Conceptual motivation and workflow placement are described in Sections \hyperref[sec:sande]{\secnameref{sec:sande}}, \hyperref[sec:candi]{\secnameref{sec:candi}}, \hyperref[sec:sands]{\secnameref{sec:sands}}, and \hyperref[sec:fandd]{\secnameref{sec:fandd}} (Figs. \ref{fig:icdssdp}–\ref{fig:finalize}). Here we provide the definitive technical account of solution evaluation, two-context execution, environment creation, catalog/collection synchronization, and interfaces required to reproduce the framework’s behavior (Fig. \ref{fig:runner-principle}; Supplementary Figs. \ref{fig:albumprinciple}–\ref{fig:api})%{We discuss the central aspects of the design principles employed in this work in isolation. For the interaction of the components, we refer to Supplementary Figure \ref{fig:albumprinciple}.}

\subsection*{Solution}
A \emph{solution} is a user-authored artifact. Album provides two framework-maintained components: (i) a \emph{controller} in the host environment that processes the solution manifest and prepares for setup and execution, and (ii) a dependency-light \emph{runner} installed in both the controller and each per-solution environment that evaluates the solution and dispatches lifecycle hooks.

\subsubsection*{Solution File}
% definition
A solution primarily consists of a single Python entry-point file. The entry-point defines (i) a machine-readable \emph{manifest} and (ii) a set of imperative lifecycle hook functions  that implement the scientific routine or their setup (see Annex \ref{supplemental:solution_example} for an example). The controller validates the manifest against a versioned JSON schema and rejects or migrates it if needed (Supplementary Fig.~\ref{fig:solution-schema}). Executable manifests provide Python-native flexibility for authors (including programmatic argument specification and hook definitions) and keep the solution artifact self-contained. However, because the controller evaluates the entry point to extract the manifest, solution authors should keep top-level code limited to manifest and hook definitions and avoid heavy imports or side effects; domain dependencies should be imported within lifecycle hooks to prevent compatibility mismatches with the controller environment. Manifest construction may be non-deterministic if it depends on external state. To mitigate this, Album validates manifests against a versioned schema and additionally supports a hybrid, directory-based solution format that externalizes selected manifest fields (Annex~\ref{supplemental:hybrid_solution}). Album does not provide sandboxing for manifest evaluation; therefore, trust relies on catalog governance and provenance.

%\chdeleted{
% hybrid approach
%Album also supports a hybrid, directory-based form of solutions, where the entry-point remains Python but selected fields of the manifest (e.g., documentation or environment specifications) may be externalized into separate files (such as Markdown or YAML) and referenced by relative paths.
% explanation
%This directory-based form additionally supports structuring larger routines by moving code into auxiliary Python modules that are imported from within lifecycle hooks.
% reason for externalizing
%Externalizing artifacts can keep the entry-point lightweight while allowing for richer documentation and backend-native environment descriptions, e.g., descriptions that do not rely on Album specific semantics.
%\todo[inline]{i should point out: during deploy they are separated for exactly this reason}
%}

\subsubsection*{Solution Schema \& Runner}
% definition
Album separates \emph{control} from \emph{execution} using a two-context architecture. 
% explanation
In the \emph{controller (host) environment}, Album evaluates the solution entry-point to obtain its manifest, validates it against the current JSON schema, and migrates older manifests to the latest schema version supported by the installed Album release (Supplementary Fig.~\ref{fig:solution-schema}). Based on the normalized manifest, the controller creates the \emph{per-solution environment} during installation, to which the execution of lifecycle hooks is then delegated.
% runner package explanation
During installation, Album installs a separate, pure-Python runner package into the solution environment at the version specified by the solution manifest (Fig.~\ref{fig:runner-principle}). This runner provides the minimal API needed to (i) reconstruct the manifest when the solution is evaluated in the solution environment, (ii) dispatch the requested lifecycle hook, and (iii) expose standardized utilities such as solution-scoped cache, download, and data paths. Because the runner is implemented in pure Python and kept dependency-light, it minimizes the risk of dependency conflicts with domain-specific packages inside the solution environment.
% migration
Schema migration is performed by the controller in the base environment to ensure that solutions authored against older schema versions remain executable under newer Album releases. The runner version pinned in the solution manifest controls the execution semantics inside the solution environment; consequently, migration is not required when a solution is executed directly inside its own environment outside of Album.
Because the solution entry-point is evaluated in the controller (host) environment, top-level code should be limited to manifest and hook definitions and should avoid heavy imports and side effects, also with regard to security. Domain dependencies should be imported within lifecycle hooks to prevent compatibility mismatches with the controller environment.

\subsection*{Environment}
% defintion
To manage \emph{per-solution environments}, Album uses micromamba\footnote{\href{https://mamba.readthedocs.io/}{https://mamba.readthedocs.io/}}, a lightweight, conda-compatible \cite{contributors_conda_2025} package manager distributed as a standalone executable. Album downloads OS/architecture-specific micromamba binaries (v2.5.0) from the official distribution endpoint. It is provided as a self-contained executable and does not require a pre-existing micromamba installation or system Python. The practical compatibility remains bounded by the host OS/kernel and architecture. 
% channels
Micromamba creates isolated environments in user space and installs packages from conda \emph{channels} (e.g., conda-forge and Bioconda). Channel coverage is platform-dependent (e.g., Bioconda does not support Windows \footnote{\href{https://bioconda.github.io/index.html}{https://bioconda.github.io/index.html}}).
% env spec
Environments are specified using the standard conda environment YAML format. In addition to conda packages, this format can include pip-installed dependencies from the Python Package Index (PyPI) via a pip, enabling access to a large ecosystem of Python libraries. 
% environment locking
For increased repeatability across operating systems, Album optionally supports dependency locking via \texttt{conda-lock}. Conda-lock performs platform-specific dependency solving and writes fully resolved lock files; micromamba can install such lock files directly, thereby reducing solver variability at install time. Locking improves repeatability by pinning dependency versions, but it does not eliminate external sources of variability (e.g., OS-level drivers or hardware constraints).

\subsection*{Catalog \& Collection}
% intro
Album's distribution layer separates \emph{shared repositories} from \emph{local state}. Solutions are authored by users and can be published to \emph{catalogs}, which are Git repositories that store both the solution artifacts and a structured index for search and browsing. Locally, each Album installation maintains a \emph{collection} that aggregates multiple catalogs and tracks local metadata and installation state. Catalogs can optionally be rendered into a static web view for human-friendly discovery, and Album supports different catalog governance modes. Indices are implemented with SQLite in our reference implementation. Additional information is available in Annex \ref{supplemental:catalog}.

\subsubsection*{Catalog}
% definition
Solutions can be gathered in catalogs, which act as shareable indices of solutions. A catalog is a Git repository that contains the solution artifacts and an accompanying SQLite database file used as an index (see Supplementary Fig.~\ref{fig:erd_catalog}). Catalogs can be stored on any filesystem (including network storage) or hosted on developer platforms such as GitLab, GitHub, or Bitbucket.
% migration
When adding a solution to a catalog or modifying an existing one, Album clones the catalog and performs a \emph{database schema migration} of the catalog's SQLite index to the latest schema version supported by the local Album installation. This allows older catalogs to remain usable, even if they were created with earlier database schema versions.
% web presence
The catalog index can be rendered into a static web view via CI/CD. For more details see Supplementary Materials, Annex \ref{supplemental:web_presence}.

\subsubsection*{Collection}
% Definition
A collection is a local aggregation of catalogs associated with an Album installation. It consists of a local SQLite database that records which catalogs are configured, indexes their solutions for search, and tracks local installation state (see Supplementary Fig.~\ref{fig:collection_schema}). Catalogs can be added to or removed from the collection. 
% migration
Catalog updates are incorporated locally by synchronizing the catalog Git repository and re-importing the updated catalog index into the local collection. Since catalog SQLite index files may evolve across Album versions, Album performs database schema migration on the catalog index before integrating its contents into the local collection, preventing schema incompatibilities and preserving usability across framework upgrades.

%Continuous Delivery/Deployment (CD) workflows. In Album, these workflows can be used to generate a static web page from a catalog by reading the catalog's SQLite index and rendering solution metadata and documentation into a browser-accessible format.
% implementation
%We implement this rendering using Gatsby{\href{https://www.gatsbyjs.com/}{https://www.gatsbyjs.com/}} and Node.js{\href{https://nodejs.org/}{https://nodejs.org/}}, and the site is updated automatically whenever the catalog index changes via the platform's CI/CD pipeline and pages hosting.}

\subsubsection*{Catalog Type}
% definition
We differentiate between \emph{direct} and \emph{indirect} catalogs. In a direct catalog, changes to the catalog (solution files and the SQLite index) are produced locally and pushed directly to the remote repository during deployment. In an indirect catalog, deployment creates a merge/pull request (MR/PR) that must be reviewed and accepted by a catalog maintainer before the catalog index is updated, enabling governance and access control.
% DOI
For indirect catalogs, CI/CD can additionally integrate with Zenodo to mint or reserve a DOI for a released solution version. Once the MR/PR is accepted, the DOI is recorded in the catalog's SQLite index, making it discoverable and queryable through the same catalog mechanisms (see Supplementary Fig.~\ref{fig:deploy}).

\subsection*{Container export}
For long-term portability, Album provides a container export mechanism that complements per-solution environments. Using the \texttt{album-docker} plugin, a solution can be packaged into an OCI-compatible container image (e.g., Docker). The plugin builds an image from an Album base image and performs an installation process, executing the solution's installation lifecycle hook, thus enabling solution-specific setup steps such as downloads or compilation, making external asset preparation part of the distributable artifact.  Building images requires access to a container runtime from the host Python process.

\subsection*{Protocols \& Interfaces}
We use the following protocols and APIs:
\begin{itemize}
    \item git transports (HTTPS/SSH/filesystem) for catalog synchronization
    \item Model Context Protocol (MCP) for large language model interaction
    \item Python API for programmatic use (Supplementary Materials, Section~\ref{supplemental:api_usage})
\end{itemize}
For internal architecture (including controllers), see Supplementary Fig.~\ref{fig:api}.
%\chdeleted{
% sync
%Catalog synchronization uses Git. Catalogs may be hosted on developer platforms and accessed via standard Git transports (HTTPS, SSH), or stored on local/network filesystems as bare repositories and accessed via Git’s filesystem transport.
% MCP
%For interaction with large language models, Album exposes its functionality via the Model Context Protocol (MCP), enabling an MCP-compatible client to discover and invoke Album operations as tools. 
% API
%In addition, Album provides a Python API for programmatic use (Supplementary Materials, Section~\ref{supplemental:api_usage}); for internal architecture (including controllers), see Supplementary Fig.~\ref{fig:api}.
%}

\section*{Code Availability}
The software suite was developed in Python and is available through PyPI and conda-forge. Additionally, we provide a standalone installer with our documentation \href{https://docs.album.solutions/}{https://docs.album.solutions/} for all major operating systems. Documentation is hosted on ReadTheDocs. Issues are tracked through GitLab \href{https://gitlab.com/album-app/album/-/issues}{https://gitlab.com/album-app/album/-/issues}. The code is open source (\href{https://gitlab.com/album-app/album}{https://gitlab.com/album-app/album}) and is published under the MIT license. We ensure the functional quality of our framework by using extensive unit/integration-testing with CI/CD.
\\

\begin{acknowledgements}
We would like to thank Iván Hidalgo, Estibaliz Gómez-de-Mariscal and Ricardo Henriques for providing the catalog \href{https://github.com/HenriquesLab/DL4MicEverywhere-album}{https://github.com/HenriquesLab/DL4MicEverywhere-album} that was used for investigating the LLM orchestration possibilities of our framework.
We further thank the Deutsche Forschungsgemeinschaft (DFG, German Research Foundation) – Project-ID 414984028 – CRC 1404 FONDA for their support. At last, we would like to express our gratitude for Jannik Franzen for his valuable feedback.
\end{acknowledgements}

\begin{contributions}
\textbf{Jan Philipp Albrecht} Conceptualisation, Investigation, Methodology, Project Administration, Software, Validation, Visualisation, Writing – Original
Draft Preparation, Writing – Review \& Editing
\textbf{Deborah Schmidt} Conceptualisation, Investigation, Methodology, Project Administration, Software, Validation, Visualisation, Writing – Original
Draft Preparation, Writing – Review \& Editing
\textbf{Maximilian Otto} Investigation, Software, Validation, Writing – Review \& Editing
\textbf{Lucas Rieckert} Software, Validation, Writing – Review \& Editing
\textbf{Kyle Harrington} Conceptualisation, Formal Analysis, Investigation, Methodology, Project Administration, Software, Supervision, Validation,
Writing – Original Draft Preparation, Writing – Review \& Editing

\end{contributions}

% literature
%\newpage
\addcontentsline{toc}{section}{References}
\bibliography{references}

\clearpage
\newpage

%\quickwordcount{}
%\quickcharcount{}
%\detailtexcount{}

%\clearpage
%\newpage

% ----------------
% --- appendix ---
% ----------------
\onecolumn 
\appendix

\lstdefinestyle{bashstyle}{
    language=bash,
    basicstyle=\ttfamily\small,
    keywordstyle=\color{red},
    deletekeywords={for,test},
    commentstyle=\color{gray},
    %stringstyle=\color{purple},
    showstringspaces=false,
    breaklines=true,
    keepspaces=true,
    upquote=true,
    morekeywords={critical,error,warning},
    sensitive=false,
    alsoletter={[]():.-_},
    literate=
    {critical}{{{\color{red}\bfseries critical}}}8
    {CRITICAL}{{{\color{red}\bfseries CRITICAL}}}8
    {error}{{{\color{red}\bfseries error}}}5
    {ERROR}{{{\color{red}\bfseries ERROR}}}5
    {warning}{{{\color{orange}\bfseries warning}}}7
    {WARNING}{{{\color{orange}\bfseries WARNING}}}7
}

\lstdefinestyle{mypython}{
  language=Python,
  basicstyle=\ttfamily\small,
  keywordstyle=\color{blue},
  commentstyle=\color{green!60!black},
  stringstyle=\color{purple},
  showstringspaces=false,
  breaklines=true
}

\section{Solution Example}
\label{supplemental:solution_example}

We exemplarily show a solution that uses three levels of exactness to define dependency resolutions: exact (pinned to patch version, e.g., python=3.14.3), approximate  (pinned to a version greater or equal to the one specified, e.g., numpy>=2.4), and free (no pinned version, e.g., matplotlib). We further equipped the solution with a basic test preparation and test execution life-cycle hook. This separation allows for altering or setting arguments specifically for testing purposes before executing the run block. When calling `album test example:solution\_example:0.1.0 --arg1 2`, the argument is first altered by the `pre\_test` block, which modifies the argument to be multiplied by 2. Its altered version is then used to evaluate the correctness of the defined scalar multiplication with the prepared matrix. Please further note that the import of `numpy` was executed within the respective life cycle hook and not at the top of the file.

Output when executed:
\begin{lstlisting}[style=bashstyle]
12:21:55 INFO    album version 0.12.1
12:21:56 INFO    ~ Starting solution_example
12:21:56 INFO    ~ My argument: 4
12:21:56 INFO    ~ Finished solution_example
12:21:56 INFO    ~ Test success!
12:21:56 INFO    Ran test routine for "solution_example"!
\end{lstlisting}

Solution file content:
\begin{lstlisting}[style=mypython]
from album.runner.api import get_args, setup

env_file = """name: example_solution
channels:
  - conda-forge
dependencies:
  - python=3.14.3
  - numpy>=2.4
  - matplotlib
"""

def album_run():
    args = get_args()
    print(f"My argument: {args.arg1}")

def album_install():
    print("Install")

def album_pre_test():
    args = get_args()
    x = args.arg1 * 2
    return {"--arg1": str(x)}

def album_test():
    import numpy as np
    args = get_args()
    assert np.all(args.arg1 * np.ones(10) == np.ones(10) * 4)
    print("Test success!")

setup(
    group="example",
    name="solution_example",
    version="0.1.0",
    album_api_version="0.7.1",
    run=album_run,
    install=album_install,
    test=album_test,
    pre_test=album_pre_test,
    dependencies={"environment_file": env_file},
    args=[
        {
            "name": "arg1",
            "description": "My integer argument.",
            "type": "integer",
            "required": True
        }
    ],
)
\end{lstlisting}

\section{Prompting}
\label{supplemental:prompting}

\lstdefinestyle{mypython}{
  language=Python,
  basicstyle=\ttfamily\small,
  keywordstyle=\color{blue},
  commentstyle=\color{green!60!black},
  stringstyle=\color{purple},
  showstringspaces=false,
  breaklines=true
}

\definecolor{mcpbg}{RGB}{247,247,249}
\definecolor{mcpframe}{RGB}{218,220,224}
\definecolor{mcpkeyword}{RGB}{0,92,175}
\definecolor{mcpcomment}{RGB}{56,130,91}
\definecolor{mcpstring}{RGB}{125,63,152}
\definecolor{mcpwrap}{RGB}{110,110,110}

\lstdefinestyle{mcpdialog}{
  basicstyle=\ttfamily\footnotesize,
  keywordstyle=\color{mcpkeyword}\bfseries,
  commentstyle=\color{mcpcomment},
  stringstyle=\color{mcpstring},
  showstringspaces=false,
  breaklines=true,
  breakatwhitespace=false,
  keepspaces=true,
  columns=fullflexible,
  upquote=true,
  frame=single,
  framerule=0.4pt,
  rulecolor=\color{mcpframe},
  backgroundcolor=\color{mcpbg},
  aboveskip=0.8\baselineskip,
  belowskip=0.8\baselineskip,
  xleftmargin=0.6em,
  framexleftmargin=0.6em,
  framextopmargin=0.5em,
  framexbottommargin=0.5em,
  tabsize=4,
  postbreak=\mbox{\textcolor{mcpwrap}{$\hookrightarrow$}\space}
}

\begin{lstlisting}[style=mcpdialog,language=Python]
    
Generate an Album solution file with the following specifications:

###album catalog: <your-catalog-name>

from album.runner.api import get_args, setup

def run():
    args = get_args()

    # Extract your arguments
    arg1 = args.arg1
    arg2 = args.arg2
    # Add more as needed

    # Your main logic here
    print(f"Running with {arg1=}, {arg2=}")
    # You can import your modules and call your pipeline here

setup(
    group="<your-group>",
    name="<your-solution-name>",
    version="0.0.1",
    title="<Title of your solution>",
    description="<Short description of what it does>",
    solution_creators=["<Your name>"],
    cite=[{"text": "<Citation text>", "url": "<Citation URL>"}],
    tags=["<tag1>", "<tag2>", "<tag3>"],
    license="MIT",
    album_api_version="0.7.1",
    args=[
        {
            "name": "arg1",
            "description": "Description for arg1",
            "type": "string",
            "required": True
        },
        {
            "name": "arg2",
            "description": "Description for arg2",
            "type": "integer",
            "required": False,
            "default": 42
        },
        # Add more arguments here
    ],
    run=run,
    dependencies={
        "parent": {
            "group": "<your-base-environment-group>",
            "name": "<your-base-environment-name>",
            "version": "<version>"
        }
    }
)

\end{lstlisting}

\newpage
\clearpage

\section{MCP usage}
\label{supplemental:mcp}

\paragraph*{Prompt:}\mbox{}\par

\begin{lstlisting}[style=mcpdialog]

    please use album to come up with a workflow for processing noisy images that need to be segmented using album solutions.

\end{lstlisting}

\paragraph*{Response:}\mbox{}\par

\begin{lstlisting}[style=mcpdialog]
    I'd be happy to create a workflow for processing noisy images that need to be segmented using Album solutions. Let me explore what's available in Album and design a comprehensive workflow.

    First, let's check what catalogs are available in Album:

    Now, let's search for solutions related to image processing, denoising, and segmentation:

    Let me try broader search terms:
    
    Now, let's search for denoising solutions:
    
    Based on the available solutions, I'll create a comprehensive workflow for processing noisy images that need to be segmented. Let me create a detailed workflow diagram and explanation.

    Based on the available solutions in the Album catalogs, I've designed a comprehensive workflow for processing noisy images that need to be segmented. The workflow addresses both 2D and 3D image processing, with separate paths for different denoising and segmentation approaches.

    ## Noisy Image Segmentation Workflow

    ### Step 1: Image Pre-processing and Denoising

    For noisy images, start with denoising to improve image quality before segmentation:

    **For 2D Images:**
    - **Supervised denoising** (when you have high-quality ground truth):
        - Use `DL4MicEverywhere:DL4MicEverywhere:care-2d-zerocostdl4mic` for supervised denoising
        - Command: `album run DL4MicEverywhere:DL4MicEverywhere:care-2d-zerocostdl4mic:1.15.2`

    - **Self-supervised denoising** (when you don't have ground truth):
        - Use `DL4MicEverywhere:DL4MicEverywhere:noise2void-2d-zerocostdl4mic` 
        - Command: `album run DL4MicEverywhere:DL4MicEverywhere:noise2void-2d-zerocostdl4mic:1.16.2`
  
        - Alternative: Use `DL4MicEverywhere:DL4MicEverywhere:deconoising-2d-zerocostdl4mic`
        - Command: `album run DL4MicEverywhere:DL4MicEverywhere:deconoising-2d-zerocostdl4mic:1.14.1`

    **For 3D Volumes:**
    - **Supervised denoising**:
        - Use `DL4MicEverywhere:DL4MicEverywhere:care-3d-zerocostdl4mic`
        - Command: `album run DL4MicEverywhere:DL4MicEverywhere:care-3d-zerocostdl4mic:1.15.3`
  
        - Alternative: Use `DL4MicEverywhere:DL4MicEverywhere:3d-rcan-zerocostdl4mic`
        - Command: `album run DL4MicEverywhere:DL4MicEverywhere:3d-rcan-zerocostdl4mic:1.14.1`

    - **Self-supervised denoising**:
        - Use `DL4MicEverywhere:DL4MicEverywhere:noise2void-3d-zerocostdl4mic`
          - Command: `album run DL4MicEverywhere:DL4MicEverywhere:noise2void-3d-zerocostdl4mic:1.16.2`

    ### Step 2: Image Segmentation

    After denoising, proceed with segmentation based on your specific needs:

    **For 2D Images:**
    - **Binary segmentation** (foreground/background):
        - Use `DL4MicEverywhere:DL4MicEverywhere:u-net-2d-zerocostdl4mic`
        - Command: `album run DL4MicEverywhere:DL4MicEverywhere:u-net-2d-zerocostdl4mic:2.2.1`

    - **Multi-label segmentation** (multiple classes):
        - Use `DL4MicEverywhere:DL4MicEverywhere:u-net-2d-multilabel-zerocostdl4mic`
        - Command: `album run DL4MicEverywhere:DL4MicEverywhere:u-net-2d-multilabel-zerocostdl4mic:2.1.4`

    - **Instance segmentation** (individual objects):
        - Use `DL4MicEverywhere:DL4MicEverywhere:stardist-2d-zerocostdl4mic` for nuclei-like objects
        - Command: `album run DL4MicEverywhere:DL4MicEverywhere:stardist-2d-zerocostdl4mic:1.20.2`
  
        - Alternative: Use `DL4MicEverywhere:DL4MicEverywhere:cellpose-2d-zerocostdl4mic` for cellular objects
        - Command: `album run DL4MicEverywhere:DL4MicEverywhere:cellpose-2d-zerocostdl4mic:1.16.3`

    - **Combined denoising and segmentation**:
        - Use `DL4MicEverywhere:DL4MicEverywhere:denoiseg-zerocostdl4mic`
        - Command: `album run DL4MicEverywhere:DL4MicEverywhere:denoiseg-zerocostdl4mic:1.14.1`

    **For 3D Volumes:**
    - **Binary segmentation**:
        - Use `DL4MicEverywhere:DL4MicEverywhere:u-net-3d-zerocostdl4mic`
        - Command: `album run DL4MicEverywhere:DL4MicEverywhere:u-net-3d-zerocostdl4mic:2.2.1`

    - **Instance segmentation**:
        - Use `DL4MicEverywhere:DL4MicEverywhere:stardist-3d-zerocostdl4mic`
        - Command: `album run DL4MicEverywhere:DL4MicEverywhere:stardist-3d-zerocostdl4mic:1.15.3`

    - **Advanced 3D segmentation** for tomographic data:
        - For tomogram datasets with complex structures, solutions from the cellcanvas catalog might be better. These include various UNet-based and Swin-UNETR models.

    ### Step 3: Post-processing and Evaluation

    After segmentation, these steps help refine and evaluate results:

    1. **For binary masks**: Use morphological operations to clean up segmentation (erosion, dilation, opening, closing)
    2. **For instance segmentation**: Separate touching objects or filter by size
    3. **Evaluation**: Compare against ground truth if available, using metrics like IoU (Intersection over Union), Dice coefficient, or F1 score
    4. **Visualization**: Display original, denoised, and segmented images side by side

    ### Step 4: Feature Extraction and Analysis

    For downstream analysis of segmented objects:

    1. **Extract metrics** like object count, size, shape, and intensity
    2. **For tomographic data**: Create 3D visualizations or extract centroids using solutions like:
        - `cellcanvas:copick:picks-from-segmentation` for extracting centroids
        - `cellcanvas:copick:voxel-counts-per-label` for quantitative analysis

    ## Sample Workflow: 3D Tomographic Data Processing

    Here's a specific example workflow for processing 3D tomographic data with noise:

    1. **Denoising**: 
   ```
   album run DL4MicEverywhere:DL4MicEverywhere:noise2void-3d-zerocostdl4mic:1.16.2
   ```

    2. **Segmentation**:
   ```
   album run DL4MicEverywhere:DL4MicEverywhere:stardist-3d-zerocostdl4mic:1.15.3
   ```

    3. **Post-processing** (for CellCanvas/Copick integration):
   ```
   album run cellcanvas:copick:paint-from-picks:0.2.3
   ```

    4. **Evaluation**:
   ```
   album run cellcanvas:copick:score-all-picks:0.0.14
   ```

    ## Best Practices

    1. **Train on representative data**: Include examples of all types of structures and noise levels you expect to encounter
    2. **Augment training data**: Especially when limited, use rotations, flips, and slight intensity variations
    3. **Use validation**: Set aside some data for validation to avoid overfitting
    4. **Test multiple approaches**: Compare different denoising and segmentation methods on a small subset before processing all data
    5. **Consider dimensionality**: While working in 3D is often ideal for volumetric data, sometimes 2D slice-by-slice processing is more practical for large datasets

    This workflow is adaptable based on your specific image characteristics, available computational resources, and the nature of the structures you need to segment.

\end{lstlisting}

\section{Hybrid solution file}
\label{supplemental:hybrid_solution}

% hybrid approach
Album also supports a hybrid, directory-based form of solutions, where the entry-point remains Python but selected fields of the manifest (e.g., documentation or environment specifications) may be externalized into separate files (such as Markdown or YAML) and referenced by relative paths.
% explanation
This directory-based form additionally supports structuring larger routines by moving code into auxiliary Python modules that are imported from within lifecycle hooks.
% reason for externalizing
Externalizing artifacts can keep the entry-point lightweight while allowing for richer documentation and backend-native environment descriptions, e.g., descriptions that do not rely on Album specific semantics.
%\todo[inline]{i should point out: during deploy they are separated for exactly this reason}

\section{Catalog and Collection index}
\label{supplemental:catalog}
Both SQLite indices are primarily a derived, query-optimized view over solution manifests. Catalog indices may also store catalog-generated metadata (e.g., DOIs minted during release workflows), which is not necessarily written back into solution files. Collection indices are built by migrating catalog entries into the collection index and store additional information such as execution information and installation status. Manifest extraction involves evaluating the solution entry-point in the controller environment to obtain the manifest and can be non-deterministic if authors compute manifest fields dynamically or if manifest values depend on environment-specific states. In practice, the determinism of catalog indexing depends on maintaining the purity of manifest construction; Album currently recommends this convention (e.g., no import-time side effects) but does not enforce it to support early collaborative interaction. Album validates the schema but does not fully sandbox manifest evaluation. Album treats catalog and collection SQLite indices as derived, query-optimized views over the authoritative solution artifacts. Indices can therefore be regenerated from solution manifests and are updated only through controlled deploy workflows (direct deploy to main or CI-generated updates after maintainer approval) to avoid merge conflicts. Solutions can be undeployed (removed from the index) even after DOI-based release; the DOI remains persistent and resolvable through Zenodo. SQLite was chosen for fast structured queries and for stable, versioned schema migration, enabling catalogs and indices to evolve alongside the framework.

\subsection*{Catalog scalability}
\label{supplemental:scalability}
We report the time required to deploy, update, and upgrade catalogs to estimate the scalability of our implementation in Supplementary Figure \ref{fig:scalability}.  Meanwhile, the catalog SQL indices require negligible disk (and repository) space consumption; however, the deployment and migration time increase noticeably with the number of registered solutions. However, usage is still possible even for large catalogs. We isolate Album’s indexing/migration overhead by measuring local operations; remote synchronization time depends on Git hosting and network and is orthogonal.

\section{Catalog Web Presence}
\label{supplemental:web_presence}
% defintition
Developer platforms such as GitLab and GitHub enable Continuous Integration (CI) and Continuous Delivery/Deployment (CD) workflows. In Album, these workflows can be used to generate a static web page from a catalog by reading the catalog's SQLite index and rendering solution metadata and documentation into a browser-accessible format.
% implementation
We implement this rendering using Gatsby\footnote{\href{https://www.gatsbyjs.com/}{https://www.gatsbyjs.com/}} and Node.js\footnote{\href{https://nodejs.org/}{https://nodejs.org/}}, and the site is updated automatically whenever the catalog index changes via the platform's CI/CD pipeline and pages hosting.

% \clearpage
% \newpage

\section{API usage and configuration}
\label{supplemental:api_usage}

Album provides an API for interaction. Calling `run()` will spawn a subprocess using the target environment’s Python, execute the "setup" routine, and call the api functionality of the pinned runner, which loads and evaluates the solution in-context and invokes the lifecycle hook. Data exchange is always file-based. Parameters can be passed to the run call as an `argv` list and must follow the semantic `--{key}={value}` or `--{key} {value}`, where key is the argument name and value is its content. `Run` returns the status of the subprocess. Log paths are configured by setting the album logger in the controller. Output can be written to a solution specific cache. The environment's working directory can be controlled in the controller via a specified path. A minimal example is available below.

\begin{lstlisting}[style=mypython]

from album.api import Album

# initialization
album_base_path = "my_album_base"
album_api = Album.Builder().base_cache_path(album_base_path).build()
album_api.load_or_create_collection()

# set solution id to work with, must be registered in collection
solution_id = "<catalog:group:name:version>"

# installation
album_api.install(solution_id)

# execution
album_api.run(solution_to_resolve=solution_id)

\end{lstlisting}

\section{Work comparison}
\label{supplemental:work_comparison}
We exemplarily work through the task of integrating two existing segmentation backends A and B with conflicting dependencies in a shared Python tool. 
Dependencies could stem, for example, from different python versions that need to be pinned for reproducibility, as shown below:

\begin{lstlisting}[style=bashstyle]]
micromamba create -n dependency_conflict -c conda-forge python=3.9 python=3.10
conda-forge/linux-64                                        Using cache
conda-forge/noarch                                          Using cache
error    libmamba Could not solve for environment specs
    The following packages are incompatible
    |-- python =3.10 * is requested and can be installed;
    '-- python =3.9 * is not installable because it conflicts with any installable versions previously reported.
critical libmamba Could not solve for environment specs
\end{lstlisting}

However, both backends should be executed from a shared Python tool from the perspective of a host-tool developer, as demonstrated in Figure \ref{fig:use-case} B. Baselines represent typical patterns rather than optimized implementations for each tool. The focus is on integration overhead induced by execution semantics and dependency isolation, not on algorithm runtime or complexity of steps.
\\
Inputs: image\\
Outputs: segmentation mask stored on disk\\
Constraint: backends A and B have dependency conflicts; both must be callable from a single host application context.
Assumption: For the Album approach, both backends are available as Album solutions. For the baselines, equivalent runnable specifications exist (environment files for manual multi-environment and Dockerfiles for containers). A minimal boundary contract for exchanging segmentation outputs (format + location) can be established. However, please note that Album does not impose a universal data interchange schema. Instead, solutions exchange artifacts via file paths and domain-appropriate conventions, which are optionally documented in metadata and validated by tests. If direct exchange is infeasible (e.g., due to versioning constraints), an explicit adapter/conversion step may be introduced (potentially as an isolated solution); this aspect is orthogonal to the dependency-isolation comparison and is not evaluated here. The necessary user rights to create environments and access the container runtime from the host process are provided. We compare the incremental integration and execution overhead within a shared host application. 

Album:
\begin{enumerate}
    \item Install Solution A
    \item Install Solution B
    \item Define and implement output exchange conventions
    \item Call A from the host Python tool
    \item Call B from the host Python tool
    \item Read outputs through standardized cache/data-path utilities (when solutions use these utilities) by following the agreed convention
\end{enumerate}
Note that `Install Solution` subsumes environment creation, dependency installation, and optional post-environment setup via the installation hook. Please further note that installation can be triggered from the host tool via the Album Python API (or CLI); however, it still creates and configures a per-solution environment; we do not compare UI/API modalities here.

Baseline A: manual multi-environment subprocess
\begin{enumerate}
    \item Create environment A from the backend A specification
    \item Install backend A
    \item Run documented post-installation steps (e.g., weights download)
    \item Create environment B from the backend B specification
    \item Install backend B
    \item Run documented post-installation steps (e.g., weights download)
    \item Define and implement output exchange conventions
    \item Run env-specific backend A from host tool (subprocess invocation)
    \item Run env-specific backend B from host tool (subprocess invocation)
    \item Read outputs by following the agreed convention
    
\end{enumerate}

Baseline B: containers
\begin{enumerate}
    \item Build image A (which includes environment post-installation steps)
    \item Build image B (which includes environment post-installation steps)
    \item Define and implement output exchange conventions
    \item Run container A (with volume mounts and GPU flags as needed) within the host tool
    \item Run container B (with volume mounts and GPU flags as needed) within the host tool
    \item Read outputs by following the agreed convention
\end{enumerate}

For a structured comparison of coordination overhead between approaches,
see Supplementary Table \ref{worktable}.

\section{Re-executability}
\renewcommand{\arraystretch}{1.5}

\label{supplemental:re_executability}
We demonstrate the backward compatibility of our framework (version 0.12.1) execution system by testing solutions with different solution\_api\_versions from python version 3.9 to 3.14. For solutions that require higher python versions than 3.9 and rely on solution\_api\_version < 0.7.0, Setuptools<81 needs to be pinned as an additional requirement. With this, all solutions, regardless of their solution\_api\_versions, remained executable under the latest framework version. We tested on the three major operating systems: Linux, Ubuntu 22.04 LTS, Windows 11 (10.0.26200), and macOS Sonoma 14.5.

\section{IO Cost}
\label{supplemental:io_cost}
Isolated execution increases I/O and storage demands. To provide an estimate regarding this demand, we report the installation time (cold vs warm cache), total disk usage added after installation, and incremental disk usage for additional solutions that share the same dependencies in figure \ref{fig:install_time}. Please note that a single container holding the same dependencies requires 4471.13 MB of disk space. This large number is due to the additional shared system dependencies necessary for the PyQT6 framework.

\newpage
\setcounter{figure}{0}
\renewcommand{\thefigure}{\arabic{figure}}
\renewcommand{\figurename}{Supplementary Figure} 

\captionsetup*{format=largeformat}

\begin{landscape}
\section{Supplementary Figures} 
\begin{figure*}[ht]
\centering
\includegraphics[width=1.4\textwidth]{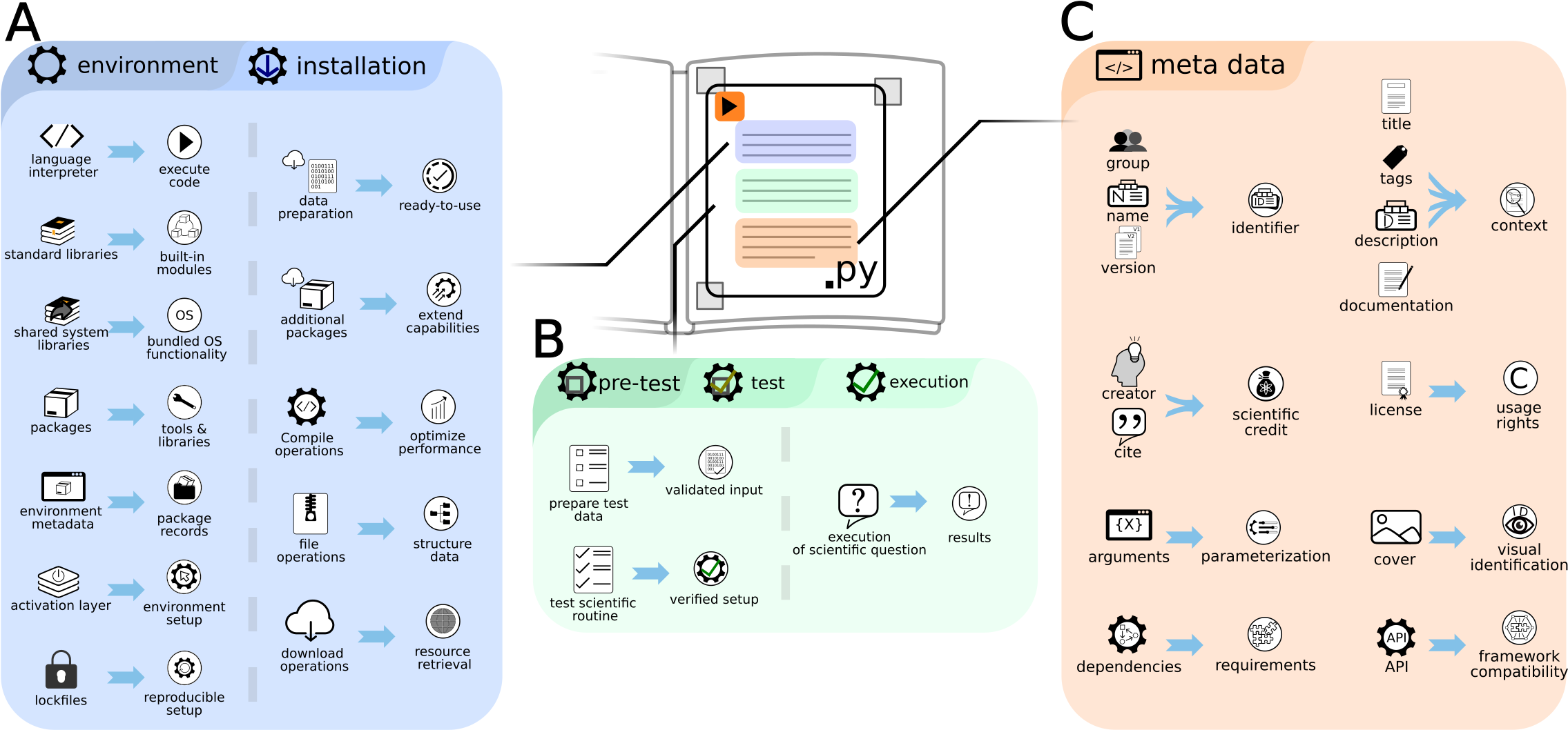}
\caption{Viewpoint of the information in a \emph{solution} file. Information is used to define three blocks: environment (\textbf{A}), execution (\textbf{B}) and meta-information (\textbf{C}). The environment block defines the runtime components and is extended by the logic provided in a post-environment function (called installation). The execution block defines the scientific routine components involving testing and scientific routine execution and the metadata block provides all necessary means for scientific context. Each component is shown with their implication for scientific work, indicated by the connection with the blue arrow.}
\label{fig:album-single-file-view}
\end{figure*}
\end{landscape}

\begin{figure*}[ht]
    \centering
    \includegraphics[width=1\linewidth]{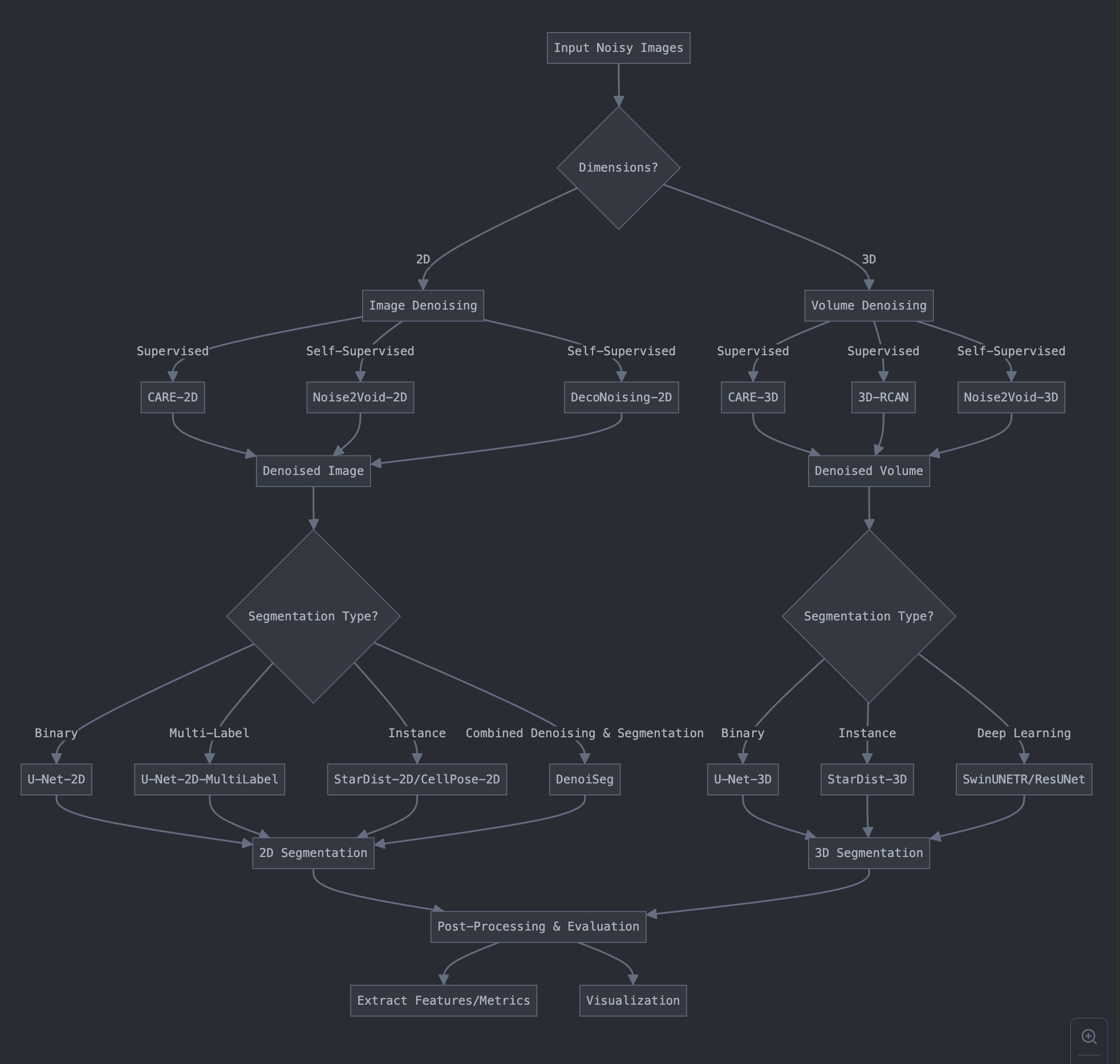}
    \caption{Example workflow generated by Claude using album MCP based on a collection with more than 500 solutions.}
    \label{fig:claude_mcp_workflow_diagram}
\end{figure*}

\begin{figure*}[ht]
    \centering
    \includegraphics[width=1\textwidth]{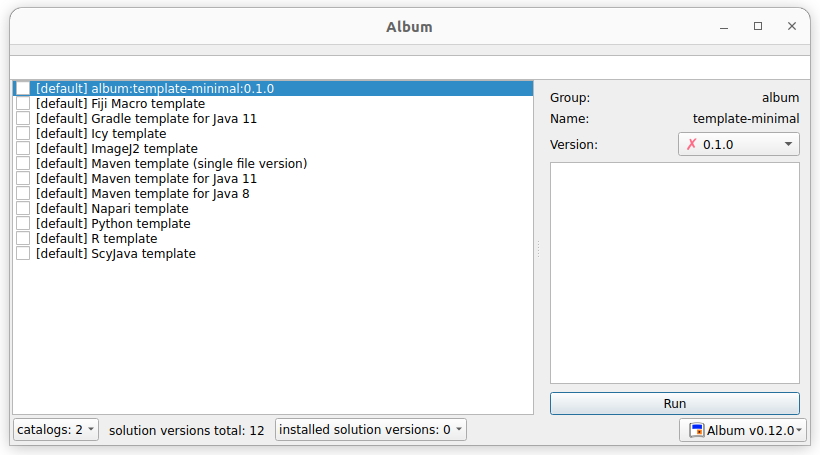}
    \caption{View on the Album-GUI start window. Visible are available solutions, as well as solution information.}
    \label{fig:album-gui}
\end{figure*}

\begin{figure*}[ht]
    \centering
    \includegraphics[width=.5\textwidth]{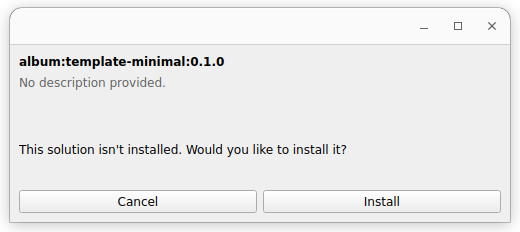}
    \caption{View on the Album-GUI solution installation window.}
    \label{fig:album-gui-install}
\end{figure*}

\begin{figure*}[ht]
    \centering
    \includegraphics[width=.5\textwidth]{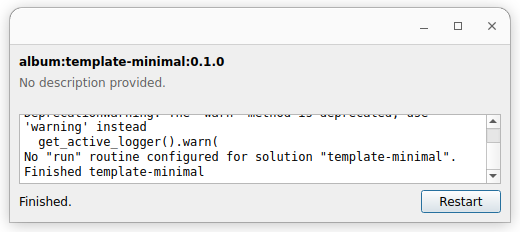}
    \caption{View on the Album-GUI solution execution window. This includes the runtime logs.}
    \label{fig:album-gui-execution}
\end{figure*}

\clearpage
\newpage

\begin{figure*}[t!]
    \centering
    \includegraphics[width=1\textwidth]{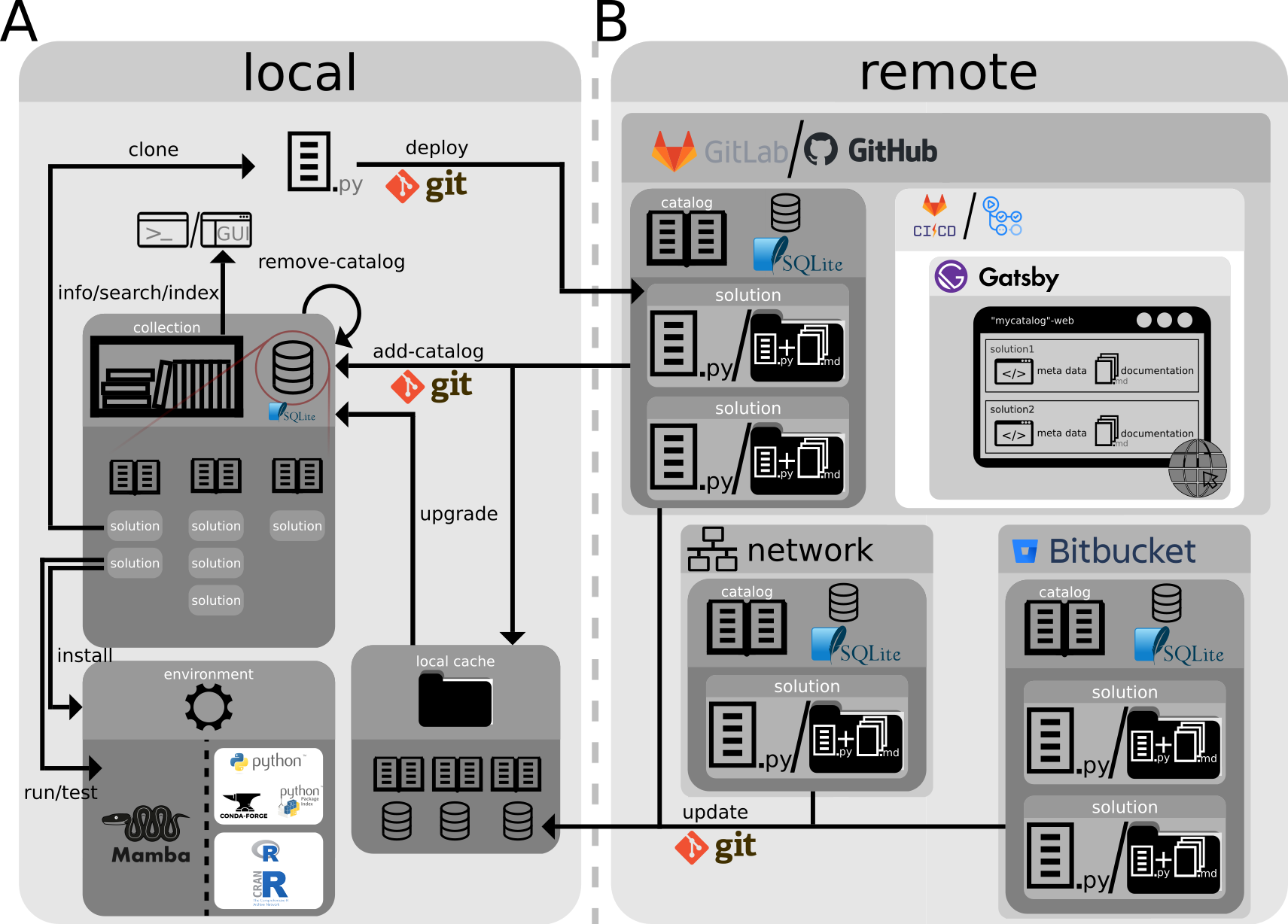}
    \caption{The design principle of the album framework. Album's design principle can be divided briefly in a part that operates locally - on the user side (\textbf{A}) - and remote - on the repository end (\textbf{B}). The software \textit{Git} is used for any command that requires communication to any remote source. Locally, an \textit{SQLite} database collects arbitrary many catalogs holding arbitrary many solutions and tracks installations, meta information and more. Environments are created for each solution during installation with \textit{Mamba}, enabling their execution within their designated environments. Exemplarily, three scenarios are shown where a catalog could exist. Please note that the web-page generation is only supported by remote sources allowing the definition of a CI/CD pipeline.}
    \label{fig:albumprinciple}
\end{figure*}

\clearpage
\newpage

% solution schema
% --- In your preamble ---
\usetikzlibrary{arrows.meta,positioning,fit,shapes.misc}

\begin{figure}[h!]
\centering
\begin{adjustbox}{max width=\textwidth, max height=\textheight}
\begin{tikzpicture}[
  node distance=10mm and 12mm,
  box/.style={
    rectangle, rounded corners=2mm, draw, align=left, inner sep=3.5pt, fill=gray!2
  },
  thinarrow/.style={-{Latex[length=2.2mm]}, line width=0.35pt},
  tinytext/.style={font=\scriptsize}
]

% -------- Central Solution object --------
\node[box, minimum width=58mm] (sol) {
  {\bfseries Solution (object)}\\[-1mm]
  \begin{tabular}{@{}l@{}}
  \texttt{group*} : string \\
  \texttt{name*} : string \\
  \texttt{version*} : string \\
  \texttt{authors} : string[] \\
  \texttt{description} : string \\
  \texttt{documentation} : string \textit{or} string[] \\
  \texttt{license} : string \\
  \texttt{tags} : string[] \\
  \texttt{title} : string \\
  \texttt{album\_api\_version} : string \\
  \texttt{changelog}, \texttt{doi} : string \\
  \texttt{args} : Argument[] \\
  \texttt{cite} : Citation[] \\
  \texttt{covers} : Cover[] \\
  \texttt{dependencies} : Dependencies \\
  \texttt{install}, \texttt{uninstall}, \texttt{run}, \texttt{test}, \texttt{pre\_test}, \texttt{close} : any
  \end{tabular}
};

% -------- Dependencies box --------
\node[box, right=of sol, xshift=8mm, minimum width=48mm] (deps) {
  {\bfseries Dependencies (object)}\\[-1mm]
  \begin{tabular}{@{}l@{}}
  \texttt{parent} : Parent \\
  \texttt{environment\_file} : string
  \end{tabular}
};

% -------- Parent BELOW Dependencies --------
\node[box, below=of deps, yshift=-2mm, minimum width=46mm] (parent) {
  {\bfseries Parent (one of)}\\[-1mm]
  \begin{tabular}{@{}l@{}}
  \textbf{A)} object with: \\
  \hspace{1.2em}\texttt{group*} : string \\
  \hspace{1.2em}\texttt{name*} : string \\
  \hspace{1.2em}\texttt{version*} : string \\[0.3em]
  \textbf{B)} object with: \texttt{resolve\_solution*} : string \\[0.3em]
  \textbf{C)} object with: \texttt{doi*} : string
  \end{tabular}
};

% -------- Other sub-objects --------
\node[box, below left=of sol, xshift=-2mm,  yshift=-12mm, minimum width=48mm] (arg) {
  {\bfseries Argument (object)}\\[-1mm]
  \begin{tabular}{@{}l@{}}
  \texttt{name*} : string \\
  \texttt{description*} : string \\
  \texttt{type} : enum\{string,file,directory,\\
  \hspace{3.25em}boolean,integer,float\} \\
  \texttt{default} : any \\
  \texttt{required} : boolean \\
  \texttt{action} : any
  \end{tabular}
};

\node[box, below=of sol, yshift=-2mm,  yshift=-12mm, minimum width=48mm] (cit) {
  {\bfseries Citation (object)}\\[-1mm]
  \begin{tabular}{@{}l@{}}
  \texttt{text*} : string \\
  \texttt{doi} : string \\
  \texttt{url} : string
  \end{tabular}
};

\node[box, below right=of sol, xshift=2mm,  yshift=-12mm, minimum width=48mm] (cov) {
  {\bfseries Cover (object)}\\[-1mm]
  \begin{tabular}{@{}l@{}}
  \texttt{source} : string \\
  \texttt{description} : string
  \end{tabular}
};

% -------- Edges --------
\draw[thinarrow] (sol.east) -- node[above, tinytext]{\texttt{dependencies}} (deps.west);
\draw[thinarrow] (sol.south west) -- node[left, tinytext, xshift=-1mm]{\texttt{args[]}} (arg.north east);
\draw[thinarrow] (sol.south) -- node[right, tinytext, xshift=1mm]{\texttt{cite[]}} (cit.north);
\draw[thinarrow] (sol.south east) -- node[left, tinytext, xshift=1mm]{\texttt{covers[]}} (cov.north west);

\draw[thinarrow] (deps.south) -- node[right, tinytext]{\texttt{parent}} (parent.north);

% -------- Legend --------
\node[below=18mm of cit] (legend) {\scriptsize\emph{Note:} Fields marked with * are required. Brackets \texttt{[]} denote arrays.};

\end{tikzpicture}
\end{adjustbox}
\caption{Diagrammatic overview of a solution schema and its related components. Please note that `album\_api\_version` is the corresponding field to map to the `album-solution-api` package.}

\label{fig:solution-schema}
\end{figure}

\clearpage
\newpage

% catalog schema
\usetikzlibrary{positioning,arrows.meta,fit,backgrounds}

\tikzset{
  table/.style={
    draw,
    rounded corners=2pt,
    align=left,
    inner sep=4pt,
    font=\sffamily\small,
    fill=white
  },
  rel/.style={-Stealth, line width=0.6pt},
  card/.style={font=\scriptsize, inner sep=1pt, fill=white, text=black}
}

\begin{figure}[h!]
\centering
\begin{adjustbox}{max width=\textheight, max height=\textwidth}

\begin{tikzpicture}[node distance=16mm and 22mm]

% ---------- CENTRAL ----------
\node[table] (solution) {
  \begin{tabular}{l}
    \textbf{solution} \\\hline
    solution\_id (PK)\\
    group \\
    name \\
    title\\
    version\\
    timestamp\\
    description\\
    doi\\
    license\\
    album\_version\\
    album\_api\_version\\
    changelog\\
    acknowledgement\\
    hash\\
  \end{tabular}
};

% ---------- SATELLITES ----------
\node[table, above left=of solution] (tag) {
  \begin{tabular}{l}
    \textbf{tag} \\\hline
    tag\_id (PK)\\
    name\\
    assignment\_type\\
  \end{tabular}
};

\node[table, left=of solution] (author) {
  \begin{tabular}{l}
    \textbf{author} \\\hline
    author\_id (PK)\\
    name\\
  \end{tabular}
};

\node[table, below left=of solution] (citation) {
  \begin{tabular}{l}
    \textbf{citation} \\\hline
    citation\_id (PK)\\
    text\\
    doi\\
    url\\
  \end{tabular}
};

\node[table, above right=of solution] (catalog) {
  \begin{tabular}{l}
    \textbf{catalog\_index} \\\hline
    name\_id (PK)\\
    name\\
    version\\
  \end{tabular}
};

% ---------- DIRECT CHILD ----------
\node[table, above=of solution] (cover) {
  \begin{tabular}{l}
    \textbf{cover} \\\hline
    cover\_id (PK)\\
    solution\_id (FK)\\
    source\\
    description\\
  \end{tabular}
};

% ---------- BRIDGE TABLES ----------
\node[table, below=of tag] (solution_tag) {
  \begin{tabular}{l}
    \textbf{solution\_tag} \\\hline
    solution\_tag\_id (PK)\\
    solution\_id (FK)\\
    tag\_id (FK)\\
  \end{tabular}
};

\node[table, below=of author] (solution_author) {
  \begin{tabular}{l}
    \textbf{solution\_author} \\\hline
    solution\_author\_id (PK)\\
    solution\_id (FK)\\
    author\_id (FK)\\
  \end{tabular}
};

\node[table, below=of solution] (solution_citation) {
  \begin{tabular}{l}
    \textbf{solution\_citation} \\\hline
    solution\_citation\_id (PK)\\
    solution\_id (FK)\\
    citation\_id (FK)\\
  \end{tabular}
};

\node[table, right=of solution] (solution_custom) {
  \begin{tabular}{l}
    \textbf{solution\_custom} \\\hline
    solution\_custom\_id (PK)\\
    solution\_id (FK)\\
    custom\_id (FK)\\
  \end{tabular}
};

\node[table, below=of solution_custom] (documentation) {
  \begin{tabular}{l}
    \textbf{documentation} \\\hline
    documentation\_id (PK)\\
    solution\_id (FK)\\
    documentation\\
  \end{tabular}
};

\node[table, below=of documentation] (solution_argument) {
  \begin{tabular}{l}
    \textbf{solution\_argument} \\\hline
    solution\_argument\_id (PK)\\
    solution\_id (FK)\\
    argument\_id (FK)\\
  \end{tabular}
};

\node[table, below=8mm of solution_argument] (argument) {
  \begin{tabular}{l}
    \textbf{argument} \\\hline
    argument\_id (PK)\\
    name\\
    type\\
    description\\
    default\_value\\
    required\\
  \end{tabular}
};

\node[table, above=10mm of solution_custom] (custom) {
  \begin{tabular}{l}
    \textbf{custom} \\\hline
    custom\_id (PK)\\
    custom\_key\\
    custom\_value\\
  \end{tabular}
};

% ---------- RELATIONSHIPS ----------
% Direct children of solution
\draw[rel] (cover) -- node[card, above left=-1pt and -1pt] {N:1} (solution);
\draw[rel] (documentation) -- node[card, below left=-1pt and -1pt] {N:1} (solution);

% Bridge relations to solution
\draw[rel] (solution_tag) -- node[card, left] {N:1} (solution);
\draw[rel] (solution_author) -- node[card, left] {N:1} (solution);
\draw[rel] (solution_citation) -- node[card, above] {N:1} (solution);
\draw[rel] (solution_argument) -- node[card, above] {N:1} (solution);
\draw[rel] (solution_custom) -- node[card, above] {N:1} (solution);

% Bridge relations to lookup tables
\draw[rel] (solution_tag) -- node[card, left] {N:1} (tag);
\draw[rel] (solution_author) -- node[card, below left] {N:1} (author);
\draw[rel] (solution_citation) -- node[card, below] {N:1} (citation);
\draw[rel] (solution_argument) -- node[card, below] {N:1} (argument);
\draw[rel] (solution_custom) -- node[card, below] {N:1} (custom);

\end{tikzpicture}

\end{adjustbox}
\caption{Logical Entity Relationship Diagram of a Catalog Index Database  showing Primary Keys (PK) and Foreign Keys (FK) as well as junction tables.}
\label{fig:erd_catalog}

\end{figure}

\clearpage
\newpage

% collection schema
\usetikzlibrary{positioning,arrows.meta,fit,backgrounds}

\tikzset{
  table/.style={
    draw,
    rounded corners=2pt,
    align=left,
    inner sep=4pt,
    font=\sffamily\small,
    fill=white
  },
  rel/.style={-Stealth, line width=0.6pt},
  card/.style={font=\scriptsize, inner sep=1pt, fill=white, text=black}
}

\begin{figure}[h!]
\centering
\begin{adjustbox}{max width=\textheight, max height=\textwidth}

\begin{tikzpicture}[node distance=16mm and 22mm]

% ---------- CENTRAL ----------

% ---------- RIGHT SIDE: CATALOG + LOOKUPS ----------
\node[table] (catalog) {
  \begin{tabular}{l}
    \textbf{catalog} \\\hline
    catalog\_id\\
    name\\
    src\\
    path\\
    branch\_name\\
    type\\
    deletable\\
  \end{tabular}
};

\node[table, left=120mm of catalog] (collection) {
  \begin{tabular}{l}
    \textbf{collection} \\\hline
    collection\_id (PK)\\
    solution\_id\\
    group\\
    name\\
    title\\
    version\\
    timestamp\\
    description\\
    doi\\
    license\\
    album\_version\\
    album\_api\_version\\
    changelog\\
    acknowledgement\\
    hash\\
    install\_date\\
    last\_execution\\
    installation\_unfinished\\
    installed\\
    catalog\_id (FK)\\
  \end{tabular}
};

\node[table, above right=35mm of collection] (documentation) {
  \begin{tabular}{l}
    \textbf{documentation} \\\hline
    documentation\_id (PK)\\
    collection\_id (FK)\\
    catalog\_id (FK)\\
    documentation\\
  \end{tabular}
};

\node[table, below=8mm of documentation] (cover) {
  \begin{tabular}{l}
    \textbf{cover} \\\hline
    cover\_id (PK)\\
    collection\_id (FK)\\
    catalog\_id (FK)\\
    source\\
    description\\
  \end{tabular}
};

\node[table, below=8mm of cover] (tag) {
  \begin{tabular}{l}
    \textbf{tag} \\\hline
    tag\_id (PK)\\
    catalog\_id (FK)\\
    name\\
    assignment\_type\\
  \end{tabular}
};

\node[table, below=8mm of tag] (author) {
  \begin{tabular}{l}
    \textbf{author} \\\hline
    author\_id (PK)\\
    catalog\_id (FK)\\
    name\\
  \end{tabular}
};

\node[table, below=8mm of author] (citation) {
  \begin{tabular}{l}
    \textbf{citation} \\\hline
    citation\_id (PK)\\
    catalog\_id (FK)\\
    text\\
    doi\\
    url\\
  \end{tabular}
};

\node[table, below=8mm of citation] (argument) {
  \begin{tabular}{l}
    \textbf{argument} \\\hline
    argument\_id (PK)\\
    catalog\_id (FK)\\
    name\\
    type\\
    description\\
    default\_value\\
    required\\
  \end{tabular}
};

\node[table, below=8mm of argument] (custom) {
  \begin{tabular}{l}
    \textbf{custom} \\\hline
    custom\_id (PK)\\
    catalog\_id (FK)\\
    custom\_key\\
    custom\_value\\
  \end{tabular}
};

% ---------- UPPER LEFT (ISOLATED/INDEX) ----------
\node[table, right=of  documentation] (catalog_collection) {
  \begin{tabular}{l}
    \textbf{catalog\_collection} \\\hline
    name\_id (PK)\\
    name\\
    version\\
  \end{tabular}
};

% ---------- RELATIONSHIPS (compact N:1 / N:N labels) ----------

% collection -> catalog
\draw[rel] (collection) -- node[card, above] {N:1} (catalog);

% lookups belong to a catalog
\draw[rel] (tag) -- node[card, above right] {N:1} (catalog);
\draw[rel] (author) -- node[card, right] {N:1} (catalog);
\draw[rel] (citation) -- node[card, below right] {N:1} (catalog);
\draw[rel] (argument) -- node[card, below] {N:1} (catalog);
\draw[rel] (custom) -- node[card, above] {N:1} (catalog);

% conceptual many-to-many between collection and each lookup (replaces collection_* junctions)
\draw[rel] (collection) -- node[card, above right] {N:N} (tag);
\draw[rel] (collection) -- node[card, right] {N:N} (author);
\draw[rel] (collection) -- node[card, below right] {N:N} (citation);
\draw[rel] (collection) -- node[card, below] {N:N} (argument);
\draw[rel] (collection) -- node[card, above] {N:N} (custom);

% cover & documentation connect to both collection and catalog
\draw[rel] (cover) -- node[card, above left] {N:1} (collection);
\draw[rel] (cover) -- node[card, above right] {N:1} (catalog);

\draw[rel] (documentation) -- node[card, below left] {N:1} (collection);
\draw[rel] (documentation) -- node[card, below right] {N:1} (catalog);

% collection self-relationship (parent-child) as N:N conceptually
\draw[rel] (collection.north west) to[bend right=50] node[card, above] {N:N (parent–child)} (collection.west);

\end{tikzpicture}

\end{adjustbox}

\caption{Conceptual Entity Relationship Diagram of a Collection Database. Primary Keys abbreviated as (PK), Foreign Keys abbreviated as (FK). Junction tables are omitted.}

\label{fig:collection_schema}

\end{figure}

\clearpage
\newpage

% catalog DOI system
\usetikzlibrary{arrows.meta,positioning,shapes.geometric,fit,calc}

\begin{landscape}

\begin{figure}[p]

\makebox[1.4\textwidth][c]{%
\begin{adjustbox}{max width=\paperwidth}

\begin{tikzpicture}[
  font=\small,
  box/.style = {draw, rounded corners, thick, align=center, minimum width=3.6cm, minimum height=1.1cm, fill=white},
  filebox/.style = {draw, rounded corners, thick, align=center, minimum width=3.6cm, minimum height=1.1cm, fill=gray!5},
  dbA/.style   = {draw, cylinder, shape border rotate=90, aspect=0.25, minimum height=1.1cm, minimum width=1.8cm, thick, fill=teal!5, align=center, draw=teal!70!black, very thick},
  db/.style    = {draw, cylinder, shape border rotate=90, aspect=0.25, minimum height=1.1cm, minimum width=1.8cm, thick, fill=white, align=center},
  swim/.style= {draw, rounded corners, inner sep=8pt, label={[anchor=north west]north west:#1}},
  arrow/.style  = {->, thick, -{Stealth[length=3mm]}},
  arrowA/.style = {->, very thick, -{Stealth[length=3.2mm]}, teal!70!black},
  arrowB/.style = {->, very thick, -{Stealth[length=3.2mm]}, blue!70!black},
  trigger/.style= {->, thick, -{Stealth[length=3mm]}, gray!70},
  triggerop/.style= {->, thick, dashed, -{Stealth[length=3mm]}, gray!70},
  stepA/.style  = {font=\footnotesize\bfseries, midway, sloped, above, text=teal!70!black},
  stepB/.style  = {font=\footnotesize\bfseries, midway, sloped, above, text=blue!70!black},
  stepBT/.style = {
  font=\footnotesize\bfseries,
  midway,
  right,
  text=blue!70!black},
  reservearrow/.style={->, very thick, dashed, -{Stealth[length=3mm]}, blue!70!black}
]

% --- Left (Local) ---
\node[box] (app) {Software\\(deploy routine)};
\node[filebox, below=7mm of app, xshift=-2.5cm] (files) {\textbf{solution files}\\(working copy)};
\node[dbA,      below=7mm of app, xshift= 2.5cm] (dblocal) {SQLite DB\\(working copy)};
\node[box, below=10mm of $(files.south)!0.5!(dblocal.south)$] (repolocal) {Git repository\\(local)};

% add an invisible dummy point 1cm below repolocal
\coordinate (dummybottom) at ($(repolocal.south)+(0,-5mm)$);

\node[
  swim,
  fit=(app)(files)(dblocal)(repolocal)(dummybottom),
  label={[yshift=2mm]above:Local (developer machine)},
  minimum width=11cm,
] (localbox) {};

% --- Center (Remote: Repo + Branches) ---
\node[box, right=7.0cm of repolocal] (feature) {origin/feature-branch};
\node[box, below=7mm of feature] (mr) {Merge Request\\(review/CI gates)};
\node[box, below=7mm of mr] (main) {origin/main};
\node[db, right=22mm of main] (dbremote) {SQLite DB};
\node[swim, fit=(feature)(mr)(main)(dbremote), label distance=2mm, 
label={[yshift=2mm]above:Remote (origin)},
minimum width=5.4cm] (remotebox) {};

% --- Right (CI/CD Runner / Platform) ---
\node[box, above=45mm of remotebox.north, anchor=south, minimum width=8cm] (ciconfig)
  {CI/CD config\\(\texttt{.yaml} in catalog repo)};

\node[box, below=7mm of ciconfig, minimum height=1.4cm] (pipeline)
  {CI/CD pipeline\\(runs on feature branch)};

\node[
swim,
fit=(ciconfig)(pipeline),
label distance=2mm,
label={[yshift=2mm]above:CI/CD runner / platform},
    minimum width=5.4cm
] (cibox){};

% --- Zenodo ---
\node[box, right=45mm of pipeline, minimum width=5cm] (zenodoreserve)
  {Reserve DOI (draft)};

\node[box, below=7mm of zenodoreserve, minimum width=5cm] (zenodopublish)
  {Release DOI \& files};

\node[
  swim,
  fit=(zenodoreserve)(zenodopublish),
  label distance=2mm,
  label={[yshift=2mm]above:Zenodo},
  minimum width=5.4cm
] (zenodobox) {};

% --- Local change flows ---
\draw[arrow] (app.west) -| node[midway, left, xshift=-2mm]{adds} (files.north);
\draw[arrowA] (app.east) -| node[midway, right, xshift= 2mm]{changes } (dblocal.north);

% --- Commit from local to repo ---
\draw[arrow] (files.south) |- node[pos=0.2, left]{git add  \textbf{solution files}} ($(repolocal.west) +(0,0.2)$);
\draw[arrowA] (dblocal.south) |- node[pos=0.2, right, text=teal!70!black]{git add DB} ($(repolocal.east)+(0,0.2)$);

% --- commit ---
\path let \p1 = (repolocal.south east) in
  coordinate (repolocalSEshift) at ($(\p1)+(-1mm,1mm)$);

\draw[->,-{Stealth[length=3mm]}]
  (repolocalSEshift)
    .. controls +(8mm,-3.2mm) and +(3.2mm,-8mm) ..
  node[right,yshift=-2mm]{git commit}
  (repolocalSEshift);

% ======================
% Path A: Direct (local DB change -> push to main)
% ======================
\draw[arrowA] (repolocal) .. controls ($(repolocal.east)!0.55!(feature.west)$) and ($(main.west)+(0,0.6)$) .. node[stepA]{push to \texttt{main}} (main.west);
\draw[arrow] (main) -- node[midway, above]{updated catalog} (dbremote);

%\node[font=\scriptsize, text=teal!70!black, above left=-2mm and -2mm of dblocal.south east, anchor=north east] {A: DB changed locally};

% ======================
% Path B: Indirect (feature -> CI alters DB -> MR -> merge)
% ======================
\draw[arrowB] (pipeline.west) -| node[stepBT, pos=0.7, left]{CI alters DB \(\rightarrow\) push} (feature.north);\draw[arrowB] (repolocal) -- node[stepB]{push feature branch (solution files only)} (feature);
\draw[trigger] ($(mr.east) +(0,-0.25)  $) -| node[pos=0.75, right]{MR triggers DB job} ($(ciconfig.south) +(2,0)   $);
\draw[arrowB] (ciconfig) -- (pipeline);
%\draw[arrowB] ($(pipeline.south) +(1.5,0) $) |- node[stepBT, pos=0.4]{CI alters DB on feature branch \(\rightarrow\) push } (mr);
\draw[arrowB] (feature) -- node[stepBT]{MR opens} (mr);
\draw[arrowB] (mr) -- node[stepBT] (mergeLabel) {admin \(\rightarrow\) merge} (main);

% Reserve DOI during feature-branch CI (B)
\draw[reservearrow] (pipeline.east) -- node[stepB, above]{reserve DOI} (zenodoreserve.west);

% After MR merge: trigger release job on main (B)
\draw[triggerop] (mergeLabel.east) -|  node[right, pos=0.6]{merge triggers release job} ($(ciconfig.south) +(3,0)   $);

% Publish DOI & upload files to Zenodo in release job
\draw[reservearrow] ($(pipeline.south) + (0,0.2) $) |- node[stepB, pos=0.7]{publish DOI \& files} (zenodopublish.west);

% --- Visual cue: in B the local DB is NOT changed ---
\node[draw=blue!40, very thick, circle, minimum size=0.6cm, inner sep=0pt, right=2mm of dblocal.east, label={[blue!70!black]right:{\scriptsize request: no local DB change}}] (nodbB) {};
\draw[blue!70!black, very thick] (nodbB.center) ++(-0.18,0.18) -- ++(0.36,-0.36);

% --- Legend ---
\coordinate (legstart) at ($(localbox.south) + (-4,-1)$);

\draw[arrowA] (legstart) -- ++(1.6,0)
  node[pos=1, right, black]{green: direct deploy routine};

\draw[arrowB] ($(legstart)+(0,-0.6)$) -- ++(1.6,0)
  node[pos=1, right, black]{blue: request deploy routine};

\draw[arrow] ($(legstart)+(0,-1.2)$) -- ++(1.6,0)
  node[pos=1, right, black]{black: applies to both, direct and deploy routines};

\draw[trigger] ($(legstart)+(0,-1.8)$) -- ++(1.6,0)
  node[pos=1, right, black]{gray: pipeline trigger};

\draw[reservearrow] ($(legstart)+(0,-2.4)$) -- ++(1.6,0)
  node[pos=1,right,black]{blue (dashed): optional request deploy routine };

\draw[triggerop] ($(legstart)+(0,-3.0)$) -- ++(1.6,0)
  node[pos=1, right, black]{gray (dashed): optional pipeline trigger};

\end{tikzpicture}
\end{adjustbox}
}

\caption{Principle behind direct and request catalogs. Request-based catalogs can optionally be configured to communicate with Zenodo. Through a two-stage merge request pipeline, DOIs can first be reserved and then, upon successful review, published. We avoid routine merge conflicts by treating the SQLite index as a deploy-time generated artifact and updating it only on the main branch (direct deploy) or in CI after maintainer approval (indirect deploy). Concurrent updates are serialized by git pushes / merge gates; failed pushes require re-sync and re-deploy.}

\label{fig:deploy}

\end{figure}

\end{landscape}

\clearpage
\newpage

% API design
\usetikzlibrary{arrows.meta,positioning,fit,shapes,calc}

\begin{landscape}

\begin{figure}[p]

\makebox[1.4\textwidth][c]{%
\begin{adjustbox}{max width=\paperwidth}

\begin{tikzpicture}[
  font=\sffamily,
  >=Latex,
  node distance=8mm and 12mm,
  block/.style={draw, rounded corners=2mm, thick, align=center, fill=white},
  iface/.style={draw, rounded corners=2mm, very thick, align=center, fill=blue!5},
  external/.style={draw, rounded corners=2mm, align=center, fill=gray!10},
  group/.style={draw, rounded corners=2mm, dashed, inner sep=6pt},
  call/.style={-Latex, thick},
  fan/.style={-Latex, semithick},
  manager/.style={draw, rounded corners=2mm, align=center, fill=green!6,
                minimum width=42mm, minimum height=12mm},
  note/.style={draw, rounded corners=1mm, fill=yellow!20, inner sep=2pt,
             font=\scriptsize, text width=42mm, align=center},
]

% USER (left)
\node[block, fill=orange!15, very thick, minimum width=32mm] (user) {User\\\small CLI / App / Script / GUI};

% ALBUM (center left)
\node[iface, very thick, minimum width=42mm, minimum height=16mm, above=22mm of user] (album) {Album\\\small public API (facade)};

% BUILDER (above Album)
\node[block, fill=purple!10, above=10mm of album, minimum width=42mm] (builder) {Album.Builder\\\small config \& build};
\node[note, above=2mm of builder] (lognote) {\strut set: base\_cache\_path, log\_format, log\_format\_time, log\_level};

\draw[call] (builder) -- node[left]{\small build()} (album);

% CONTROLLER (center)
\node[block, fill=blue!10, minimum width=42mm, minimum height=16mm, right=22mm of album] (controller) {AlbumController\\\small internal orchestration};

\draw[call] (user) -- node[right]{\small e.g.\ run(), install()} (album);
\draw[call] (album) -- node[above]{\small delegate} (controller);

% === MANAGERS GRID ===
% Row 1 (wider column gaps)
%\node[manager, right=26mm of controller] (eventm) {Event\\Manager};
\node[manager, right=26mm of controller] (deploym) {Deploy\\Manager};
\node[manager, right=18mm of deploym] (catalogs) {Catalogs\\(view)};
% Row 2 (taller row gap)
\node[manager, below=12mm of deploym] (installm) {Install\\Manager};
\node[manager, below=12mm of catalogs] (searchm) {Search\\Manager};
\node[manager, right=18mm of searchm] (statem) {State\\Manager};
% Row 3
\node[manager, below=12mm of installm] (testm) {Test\\Manager};
\node[manager, below=12mm of searchm] (runm) {Run\\Manager};
\node[manager, below=12mm of statem] (collection) {Collection\\Manager};
% Row 4
\node[manager, below=12mm of testm] (taskm) {Task\\Manager};
\node[manager, below=12mm of runm] (scriptm) {Script\\Manager};
\node[manager, below=12mm of collection] (migrationm) {Migration\\Manager};

% === EXTERNAL SYSTEMS (placed under three columns to avoid anchor tricks) ===
\node[external, below=8mm of migrationm, xshift=-28mm] (fs) {Local Collection \& Index\\\small (FS/DB)};
\node[external, below=8mm of taskm, xshift=28mm] (envs) {Solution Environments\\\small (venv/conda, downloads)};
\node[external, above=18mm of deploym, xshift=-28mm] (git) {Catalog Repos\\\small (Git/Remote)};
\node[external, left=12mm of taskm] (bus) {Event Bus /\\Task Queue};

% Manager -> external relations
\draw[call] (collection.west) -| (fs.north);
\draw[call] (migrationm.west) -| (fs.north);
\draw[call] (installm.east) -| (envs.north);
\draw[call] (runm.west) -| (envs.north);
\draw[call] (deploym.west) -| (git.south);
%\draw[call] (eventm.east) |- (bus.north);
\draw[call] (taskm.west) -- (bus.east);

% Method hints directly above each manager
\node[note, above=2mm of collection.north] {\scriptsize load\_or\_create(), get\_index\_as\_dict(), resolve()};
\node[note, above=2mm of catalogs.north] (catalogstext) {\scriptsize get\_catalog\_by\_name/src(), get\_catalogs\_as\_dict()};
\node[note, above=2mm of runm.north] {\scriptsize run(), run\_solution\_script()};
\node[note, above=2mm of installm.north] {\scriptsize install(), uninstall(), is\_installed()};
\node[note, above=2mm of searchm.north] {\scriptsize search(keywords)};
\node[note, above=2mm of statem.north] {\scriptsize load(path)};
\node[note, above=2mm of testm.north] {\scriptsize test(solution)};
\node[note, above=2mm of deploym.north] (deploytext) {\scriptsize deploy(), undeploy(), clone()};
\node[note, above=2mm of scriptm.north] {\scriptsize script hooks};
%\node[note, above=2mm of eventm.north] {\scriptsize add/remove listener, publish\_event()};
\node[note, above=2mm of taskm.north] {\scriptsize \_run\_async(), create\_and\_register\_task(), get\_task\_status(), finish\_tasks()};
\node[note, above=2mm of migrationm.north] {\scriptsize load\_catalog\_index()};

% Group box (purely visual)
\node[group, fit=(collection) (runm) (catalogs) (installm) (searchm) (statem) (testm) (deploym) (scriptm) (taskm) (deploytext) (catalogstext) (migrationm),
      label={[anchor=north west]north west:\small Managers}] (mgrbox) {};

% Single arrow from Controller to Manager box
\draw[call] (controller.east) -- (mgrbox.west) node[midway,above,sloped]{\small delegates to managers};

% Logger lifecycle hint
\node[note, left=-10mm of controller, align=left, yshift=13mm] (log)
  {\scriptsize configure\_root\_logger() on build\\\scriptsize pop\_active\_logger() on close()};

% Simple legend (no tabular/minipage)
\node[group, below=56mm of album, inner sep=4pt, align=left] (legend)
  {\footnotesize
   \shortstack[l]{%
     \textbf{Legend}\\
     \textcolor{black}{\rule{7mm}{3mm}} Call path (arrow)\\
     \textcolor{blue!5}{\rule{7mm}{3mm}} Album (public API)\\
     \textcolor{blue!10}{\rule{7mm}{3mm}} Controller (internal)\\
     \textcolor{green!6}{\rule{7mm}{3mm}} Managers\\
     \textcolor{gray!10}{\rule{7mm}{3mm}} External systems
   }%
  };

\end{tikzpicture}

\end{adjustbox}
}

\caption{Public API design with internal controller and managers. }

\label{fig:api}

\end{figure}

\end{landscape}

\clearpage
\newpage

% install time
\begin{figure*}[ht]
    \centering
    \includegraphics[width=1\linewidth]{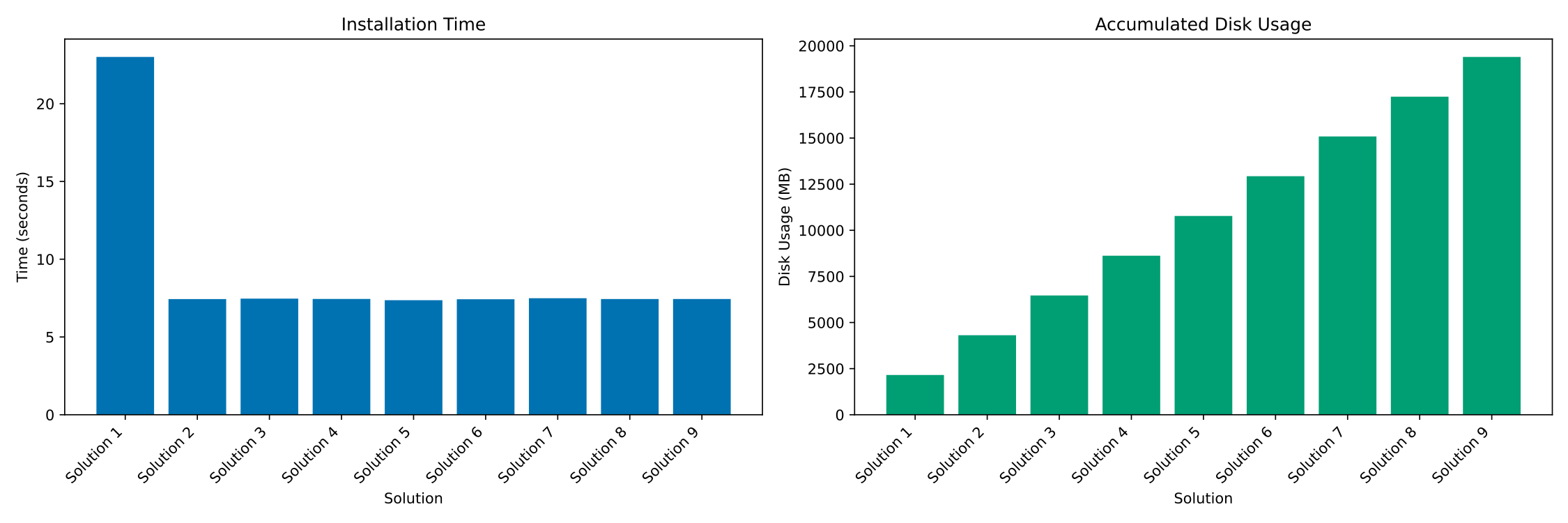}
    \caption{Installation Time and Disk usage: Left shows the installation time of solutions with identical dependencies (Python and PyQT6). First installation uses cold cache, the following installations benefit from caching. Right shows analogously the accumulated disk usage for each isolated environment created during installation.}
    \label{fig:install_time}
\end{figure*}

% scalability
\begin{figure*}[ht]
    \centering
    \includegraphics[width=1\linewidth]{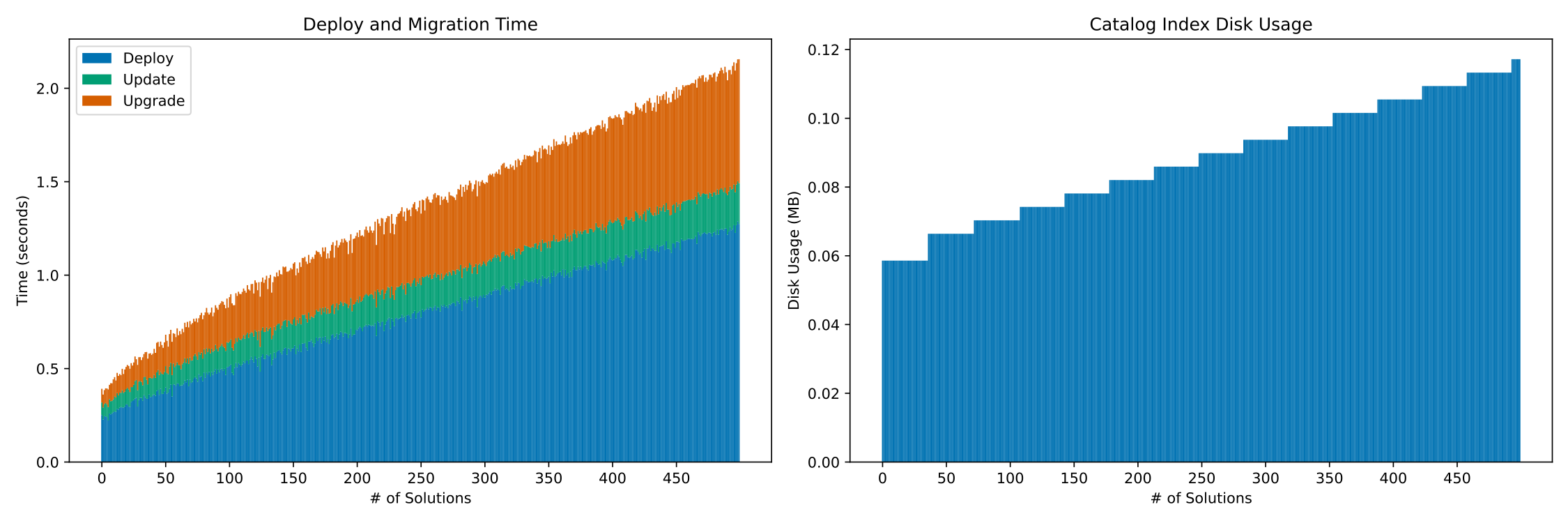}
    \caption{Catalog and Index scalability. Left shows the Deploy time and corresponding update (git fetch) and upgrade (migration into collection) time for a growing number of solutions. The right plot shows the corresponding catalog disk growth. The collection index grows analogously and is therefore not shown.}
    \label{fig:scalability}
\end{figure*}

% Text %
%%%%%%%%

% Tables %
%%%%%%%%%%%
\begin{landscape}

\section{Tables}
\setcounter{table}{0}
\renewcommand{\thetable}{\arabic{table}}
\renewcommand{\tablename}{Supplementary Table} 

\begin{table*}[ht!]

\centering
\begin{adjustbox}{}%angle=90}
\resizebox{1.4\textwidth}{!}{%

\begin{tabular}{c||*{3}{c|}|*{3}{c|}|*{3}{c|}|*{4}{c|}|}
    \hline
    & \multicolumn{3}{|c||}{\makecell[l]{A}} & \multicolumn{3}{c||}{\makecell[l]{B}} & \multicolumn{3}{c||}{\makecell[l]{C}} & \multicolumn{4}{c||}{\makecell[l]{D}} \\
    & \multicolumn{3}{|c||}{Search \& Explore} & \multicolumn{3}{c||}{Creation \& Implementation} & \multicolumn{3}{c||}{Share \& Scale} & \multicolumn{4}{c||}{Finalize \& Distribute} \\
    \hline
    \hline
    
\rotatebox[origin=c]{90}{ category } & 
\makecell{\raisebox{-1mm}{\includegraphics[width=0.05\textwidth]{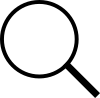}} \\ find} & 
\makecell{\raisebox{-1mm}{\includegraphics[width=0.05\textwidth]{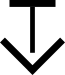}} \\ access} & 
\makecell{\raisebox{-1mm}{\includegraphics[width=0.05\textwidth]{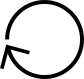}} \\ try} & 
\makecell{\raisebox{-1mm}{\includegraphics[width=0.05\textwidth]{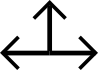}} \\ adapt} & 
\makecell{\raisebox{-1mm}{\includegraphics[width=0.05\textwidth]{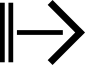}} \\ extend} & 
\makecell{\raisebox{-1mm}{\includegraphics[width=0.05\textwidth]{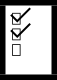}} \\ test} & 
\makecell{\raisebox{-1mm}{\includegraphics[width=0.05\textwidth]{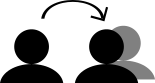}} \\ share} & 
\makecell{\raisebox{-1mm}{\includegraphics[width=0.05\textwidth]{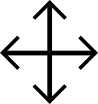}} \\ scale} & 
\makecell{\raisebox{-1mm}{\includegraphics[width=0.05\textwidth]{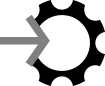}} \\ integrate} & 
\makecell{\raisebox{-1mm}{\includegraphics[width=0.05\textwidth]{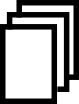}} \\ document} & 
\makecell{\raisebox{-1mm}{\includegraphics[width=0.05\textwidth]{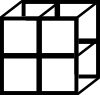}} \\ package} & 
\makecell{\raisebox{-1mm}{\includegraphics[width=0.05\textwidth]{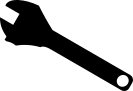}} \\ maintain} & 
\makecell{\raisebox{-1mm}{\includegraphics[width=0.05\textwidth]{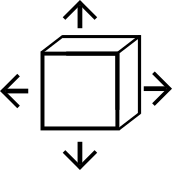}} \\ export} \\
    \hline

    \rotatebox[origin=c]{90}{ commandline } & \makecell[l]{>add\_catalog\\>search\\>index} & \makecell[l]{>info\\>install \\ \quad} & \makecell[l]{>run \\ >test \\ \quad} & \makecell[l]{>repl \\ \quad \\ \quad } & \makecell[l]{>clone \\ \quad \\ \quad} & \makecell[l]{>test \\ \quad \\ \quad} & \makecell[l]{>deploy \\ >upgrade \\ >update }  & \makecell[l]{>run-batch \\ \quad \\ \quad} & \makecell[l]{} & \makecell[l]{ } & \makecell[l]{>containerize \\ \quad \\ \quad} & \makecell[l]{>deploy \\ \quad \\ \quad} & \makecell[l]{>docker \\ \quad \\ \quad} \\
    \hline
    
    \rotatebox[origin=c]{90}{ keywords } & \makecell[l]{add catalogs \\ search in catalogs \\ list solutions } & \makecell[l]{install solutions \\ list information \\ \quad} & \makecell[l]{run solutions\\test solutions \\ \quad} & \makecell[l]{interactive console\\debugging\\experimenting} & \makecell[l]{clone \\ prototyping \\ re-iterate solutions} & \makecell[l]{quality control\\ \quad \\ \quad} & \makecell[l]{collaborate \\ distribute \\ enable usage} & \makecell[l]{batch run \\ looping \\ \quad} & \makecell[l]{integrate in code \\ escape version conflicts \\ \quad} & \makecell[l]{document usage \\ \quad \\ \quad} & \makecell[l]{container \\ \quad} & \makecell[l]{increase version \\ change-log \\ \quad} & \makecell[l]{finalize \\ \quad \\ \quad} \\
    \hline

    \rotatebox[origin=c]{90}{ GUI } & yes & yes & yes & no & no & no & no & no & no & no & no & no & no \\
    \hline
    
    \rotatebox[origin=c]{90}{ module } & core & core & core & core & core & core & core & plugin & core & core & plugin & core & plugin\\
    \hline
    
    \rotatebox[origin=c]{90}{ method } & \makecell[l]{gatsby \\ git \\ sqlite} & \makecell[l]{git \\ sqlite} & \makecell[l]{mamba} & \makecell[l]{mamba} & \makecell[l]{git} & \makecell[l]{mamba} & \makecell[l]{git \\ gatsby} & \makecell[l]{mamba} & \makecell[l]{python API} & \makecell[l]{markdown \\ gatsby} & \makecell[l]{pyinstaller} & \makecell[l]{git \\ sqlite} & \makecell[l]{docker}\\
    \hline
    
    \rotatebox[origin=c]{90}{ group } & \makecell[l]{user} & \makecell[l]{user} & \makecell[l]{user\\developer} & \makecell[l]{developer} & \makecell[l]{developer} & \makecell[l]{developer} & \makecell[l]{developer} & \makecell[l]{user\\developer} & \makecell[l]{developer} & \makecell[l]{user} & \makecell[l]{user\\developer} & \makecell[l]{developer} & \makecell[l]{developer}\\
    \hline
\end{tabular}
}
\end{adjustbox}
\caption{Album clue card. The rows show the challenge category, command-line calls, availability of a GUI view, underlying methodology, modularity and target group. The columns stand for the four fields of the iterative, collaboration-driven research software cycle (A to D) and their corresponding technical challenge.}
\label{tab:album-clue-card}
\end{table*}

\end{landscape}

\newpage
\clearpage

\renewcommand{\thetable}{\arabic{table}}
\renewcommand{\tablename}{Supplementary Table} 

\newcolumntype{L}[1]{>{\raggedright\arraybackslash}p{#1}}
\begin{table*}[ht]
    \centering
    
    \resizebox{.8\textwidth}{!}{%
    \small
        \begin{tabular}{p{.75cm}|p{8cm}|p{1cm}|L{8cm}}
        & Definition & \multicolumn{2}{c}{Condition} \\
        \hline
        \hline
        F & \multicolumn{3}{p{16cm}}{Software, and its associated metadata, is easy for both humans and machines to find. } \\ \hline
        F1. & Software is assigned a globally unique and persistent identifier. & \includegraphics[width=0.03\textwidth]{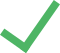} & Yes, when using a catalog with Zenodo support, a DOI can be reserved for a solution per deploy (versioned release).\\
        F1.1. & Components of the software representing levels of granularity are assigned distinct identifiers. & \includegraphics[width=0.03\textwidth]{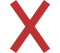} & Inherently, no persistent IDs for sub-components. However solutions can be used and assembled in a modular manner, and as such can have distinct DOIs.\\
        F1.2. & Different versions of the software are assigned distinct identifiers. & \includegraphics[width=0.03\textwidth]{figures/figure_icon/check.png} & Yes, a Zenodo DOI can be attached to a solution per deploy (release). Zenodo gives both a concept DOI and version DOIs.\\
        F2. & Software is described with rich metadata. & \includegraphics[width=0.03\textwidth]{figures/figure_icon/check.png} & Yes, our json metadata schema includes fields such as group, name, version, DOI, Authors, documentation, description, tags, change-log , acknowledgments. \\
        F3. & Metadata clearly and explicitly include the identifier of the software they describe. & \includegraphics[width=0.03\textwidth]{figures/figure_icon/check.png} & Yes, the metadata of the software includes a DOI. \\
        F4. & Metadata are FAIR, searchable and indexable. & \includegraphics[width=0.03\textwidth]{figures/figure_icon/cross.png} & FAIR harvesting is possible via Zenodo, however this does not fully comply with FAIR criteria. Therefore, registering software to repositories such as Helmholtz Research Software Directories is recommended. Album catalogs do not expose metadata to global harvesters by itself.\\ \hline
        
        A & \multicolumn{3}{p{16cm}}{A: Software, and its metadata, is retrievable via standardised protocols.} \\ \hline
        A1. & Software is retrievable by its identifier using a standardised communications protocol. & \includegraphics[width=0.03\textwidth]{figures/figure_icon/check.png} & Yes, when catalogs are configured with Zenodo DOI support, the software can be retrieved by specifying the DOI.\\
        A1.1. & The protocol is open, free, and universally implementable. & \includegraphics[width=0.03\textwidth]{figures/figure_icon/check.png} & Yes, with album supporting git ssh, HTTPS, and the Git protocol. \\ 
        A1.2. & The protocol allows for an authentication and authorization procedure, where necessary. & \includegraphics[width=0.03\textwidth]{figures/figure_icon/check.png} & Yes, with album authentication procedure is possible. Authentication via SSH keys/Personal Access Tokens; Zenodo via API token.\\
        A2. & Metadata are accessible, even when the software is no longer available. & \includegraphics[width=0.03\textwidth]{figures/figure_icon/check.png} & Yes, with connecting solutions to a DOI metadata records persist in Zenodo even when the repository is deleted. Therefore, persistence is guaranteed by Zenodo, not by an Album catalog.\\ \hline
        
        I & \multicolumn{3}{p{16cm}}{Software interoperates with other software by exchanging data and/or metadata, and/or through interaction via application programming interfaces (APIs), described through standards. } \\ \hline
        I1. & Software reads, writes and exchanges data in a way that meets domain-relevant community standards. & \includegraphics[width=0.03\textwidth]{figures/figure_icon/cross.png} & Depends on the solution implementation. Cannot be covered by Album. Templates can encourage use of community standards, but Album does not enforce them. \\
        I2. & Software includes qualified references to other objects. & \includegraphics[width=0.03\textwidth]{figures/figure_icon/cross.png} & Our schema provides means to connect the solution to a publication or other resource by citation. However these are not machine-actionable yet. \\ \hline
        
        R & \multicolumn{3}{p{16cm}}{Software is both usable (can be executed) and reusable (can be understood, modified, built upon, or incorporated into other software). } \\ \hline
        R1. & Software is described with a plurality of accurate and relevant attributes. & \includegraphics[width=0.03\textwidth]{figures/figure_icon/check.png} & Yes, authors can use tags, provide a description, and attach a documentation.\\
        R1.1. & Software is given a clear and accessible license. & \includegraphics[width=0.03\textwidth]{figures/figure_icon/check.png} & Yes, authors are encouraged to include a license together with their solution. Album provides license field and default, but compliance depends on authors. \\
        R1.2. & Software is associated with detailed provenance. & \includegraphics[width=0.03\textwidth]{figures/figure_icon/check.png} & Yes, the schema of a solution consists of: group, name, version, DOI, Authors, documentation, description, tags, change-log, acknowledgments, environment (e.g. dependenies). When deployed to a catalog with Zenodo support, metadata records are created. \\
        R2. & Software includes qualified references to other software. & \includegraphics[width=0.03\textwidth]{figures/figure_icon/check.png} & Yes, references can be defined by providing a DOI, supported at descriptive level; not yet machine-actionable.\\
        R3. & Software meets domain-relevant community standards. & \includegraphics[width=0.03\textwidth]{figures/figure_icon/cross.png} & Depends on the solution implementation. Cannot be covered by Album.\\

    \multicolumn{4}{c}{\makecell[l]{  \includegraphics[width=0.03\textwidth]{figures/figure_icon/check.png}  \quad Framework provides capability. Author responsible for implementation and optimal framework configuration. }} \\
    \multicolumn{4}{c}{\makecell[l]{  \includegraphics[width=0.03\textwidth]{figures/figure_icon/cross.png}  \quad Framework does not provide capability. Feature depends on solution implementation context. }} \\
        
        \end{tabular}
    }
    \caption{FAIR-enabling features of Album (and conditions on solutions).}
    \label{tab:fair4rsalbum}
\end{table*}

\clearpage
\newpage

\begin{landscape}
\begin{table*}

\centering
\begin{adjustbox}{ }%angle=90}

\resizebox{1.4\textwidth}{!}{%

\begin{tabular}{|l|l|*{12}{c|}}
\hline
\multicolumn{2}{|l|}{\diagbox{Category + Feature}{Tool}} & our tool & UV & mamba & pixie & poetry & Guix & nextflow & snakemake & knime (desktop) & galaxy & docker & apptainer \\
\hline

% %%%%%%%%%%%%%%% Ecosystem %%%%%%%%%%%%%%%
\multirow{4}{*}{\rotatebox[origin=c]{90}{Ecosystem}}

% How does the tool define itself?
& description                   & \makecell[c]{ - } & \makecell[c]{package manager \\ project manager} & \makecell[c]{package manager} & \makecell[c]{package management} & \makecell[c]{project manager \\ package management} & \makecell[c]{functional \\ package management}  &  \makecell[c]{ workflow management}  &  \makecell[c]{workflow management} &  \makecell[c]{workflow management} &  \makecell[c]{workflow management} &  \makecell[c]{container engine} & \makecell[c]{container engine} \\

% What primary model for getting dependencies or defintions of the tool is used?
& \makecell[l]{source model or \\ dependency type}       & \makecell[c]{Open Registry (PyPI) \\ Named channels} & \makecell[c]{Open Registry (PyPI)} & \makecell[c]{Open Registry (PyPI) \\ Named channels}  &  \makecell[c]{Open Registry (PyPI) \\ Named channels}  & \makecell[c]{Open Registry (PyPI) \\ Closed Registry} & \makecell[c]{Built-In Registry} & \makecell[c]{Package managers \\ Container runtimes \\ Dynamic provisioning } & \makecell[c]{Package managers \\ Container runtimes} & \makecell[c]{Open Registry \\ (KNIME Hub)}  & \makecell[c]{Package managers \\ Container runtimes} & \makecell[c]{Open Registry \\(Docker Hub)} & \makecell[c]{Open Registry \\(Docker Hub)} \\

% Is the code available and open source?
& open source  &  \includegraphics[width=0.03\textwidth]{figures/figure_icon/check.png}   &  \includegraphics[width=0.03\textwidth]{figures/figure_icon/check.png}   &  \includegraphics[width=0.03\textwidth]{figures/figure_icon/check.png}  &  \includegraphics[width=0.03\textwidth]{figures/figure_icon/check.png}   &  \includegraphics[width=0.03\textwidth]{figures/figure_icon/check.png}  &  \includegraphics[width=0.03\textwidth]{figures/figure_icon/check.png}  &  \includegraphics[width=0.03\textwidth]{figures/figure_icon/check.png}  &  \includegraphics[width=0.03\textwidth]{figures/figure_icon/check.png}  &  \includegraphics[width=0.03\textwidth]{figures/figure_icon/check.png}   &  \includegraphics[width=0.03\textwidth]{figures/figure_icon/check.png}    &  \includegraphics[width=0.03\textwidth]{figures/figure_icon/cross.png}  &  \includegraphics[width=0.03\textwidth]{figures/figure_icon/check.png}  \\

% What license do the tool have?
& license & MIT & \makecell[c]{ Apache-2.0 \\ MIT}  & BSD-3-Clause & BSD-3-Clause  & MIT &  GPLv3  & Apache-2.0 & MIT & GPLv3 (mostly) & MIT & Apache-2.0 & BSD-3-Clause \\

\hline

% %%%%%%%%%%%%% Reproducibility & Transparency %%%%%%%%%%%%%
\multirow{5}{*}{\rotatebox[origin=c]{90}{\footnotesize Reproducibility / Transparency}}

% is there a way to store the result of the tool in a reproducible portable way?
& Supports portability                     &  \includegraphics[width=0.03\textwidth]{figures/figure_icon/check.png}  &  \includegraphics[width=0.03\textwidth]{figures/figure_icon/check.png}  &  \includegraphics[width=0.03\textwidth]{figures/figure_icon/check.png}  &  \includegraphics[width=0.03\textwidth]{figures/figure_icon/check.png}  &  \includegraphics[width=0.03\textwidth]{figures/figure_icon/check.png}  &  \includegraphics[width=0.03\textwidth]{figures/figure_icon/check.png}  &  \includegraphics[width=0.03\textwidth]{figures/figure_icon/check.png}  &  \includegraphics[width=0.03\textwidth]{figures/figure_icon/check.png}  &  \includegraphics[width=0.03\textwidth]{figures/figure_icon/cross.png}   &  \includegraphics[width=0.03\textwidth]{figures/figure_icon/check.png}  &  \includegraphics[width=0.03\textwidth]{figures/figure_icon/check.png}  &  \includegraphics[width=0.03\textwidth]{figures/figure_icon/check.png}  \\

% What format is used to enable reproducibility?
& \makecell[l]{reproducible definition format}    & \makecell[c]{Python \\ (YAML)} & \makecell[c]{Python \\ TOML} & \makecell[c]{YAML} & \makecell[c]{TOML} & \makecell[c]{TOML} & Scheme & groovy & Python & GUI & GUI & Dockerfile & Apptainer Definition File \\

% Can users export the outcome of the tool to a container?
& export to container \textsuperscript{\dag}     &  \includegraphics[width=0.03\textwidth]{figures/figure_icon/cross.png}  &  \includegraphics[width=0.03\textwidth]{figures/figure_icon/cross.png}  &  \includegraphics[width=0.03\textwidth]{figures/figure_icon/cross.png}  &  \includegraphics[width=0.03\textwidth]{figures/figure_icon/cross.png}  &  \includegraphics[width=0.03\textwidth]{figures/figure_icon/cross.png}  &  \includegraphics[width=0.03\textwidth]{figures/figure_icon/check.png}  &  \includegraphics[width=0.03\textwidth]{figures/figure_icon/cross.png}  &  \includegraphics[width=0.03\textwidth]{figures/figure_icon/cross.png}  &  \includegraphics[width=0.03\textwidth]{figures/figure_icon/cross.png}   &  \includegraphics[width=0.03\textwidth]{figures/figure_icon/cross.png}  &  \includegraphics[width=0.03\textwidth]{figures/figure_icon/check.png}  &  \includegraphics[width=0.03\textwidth]{figures/figure_icon/check.png}  \\

% Does the tool record and preserve structured metadata about executed steps, parameters, tools, versions, and environments in a way that supports reproducibility or auditability over time?
& provenance tracking         &  \includegraphics[width=0.03\textwidth]{figures/figure_icon/cross.png}   &  \includegraphics[width=0.03\textwidth]{figures/figure_icon/cross.png}  &  \includegraphics[width=0.03\textwidth]{figures/figure_icon/cross.png}  &  \includegraphics[width=0.03\textwidth]{figures/figure_icon/cross.png}  &  \includegraphics[width=0.03\textwidth]{figures/figure_icon/cross.png}   &  \includegraphics[width=0.03\textwidth]{figures/figure_icon/check.png}  &  \includegraphics[width=0.03\textwidth]{figures/figure_icon/check.png}   &  \includegraphics[width=0.03\textwidth]{figures/figure_icon/check.png}  &  \includegraphics[width=0.03\textwidth]{figures/figure_icon/cross.png}  &  \includegraphics[width=0.03\textwidth]{figures/figure_icon/check.png}  &  \includegraphics[width=0.03\textwidth]{figures/figure_icon/cross.png}  &  \includegraphics[width=0.03\textwidth]{figures/figure_icon/cross.png}  \\

% Can the tool export dependencies of scientific work?
& inspectable dependency metadata \textsuperscript{\P}         &  \includegraphics[width=0.03\textwidth]{figures/figure_icon/check.png}  &  \includegraphics[width=0.03\textwidth]{figures/figure_icon/check.png}  &  \includegraphics[width=0.03\textwidth]{figures/figure_icon/check.png}  &  \includegraphics[width=0.03\textwidth]{figures/figure_icon/check.png}  &  \includegraphics[width=0.03\textwidth]{figures/figure_icon/check.png}  &  \includegraphics[width=0.03\textwidth]{figures/figure_icon/check.png}  &  \includegraphics[width=0.03\textwidth]{figures/figure_icon/cross.png}  &  \includegraphics[width=0.03\textwidth]{figures/figure_icon/cross.png}  &  \includegraphics[width=0.03\textwidth]{figures/figure_icon/cross.png}  &  \includegraphics[width=0.03\textwidth]{figures/figure_icon/cross.png}  &  \includegraphics[width=0.03\textwidth]{figures/figure_icon/cross.png}  &  \includegraphics[width=0.03\textwidth]{figures/figure_icon/cross.png}  \\

% Does the tool support assigning or integrating with DOI systems for workflows, environments, packages, or containers — either directly or via integration (e.g., Zenodo)?
%& DOI support for artifacts &  \includegraphics[width=0.03\textwidth]{figures/figure_icon/check.png}  &  \includegraphics[width=0.03\textwidth]{figures/figure_icon/cross.png}  &  \includegraphics[width=0.03\textwidth]{figures/figure_icon/cross.png}  &  \includegraphics[width=0.03\textwidth]{figures/figure_icon/cross.png}  &  \includegraphics[width=0.03\textwidth]{figures/figure_icon/cross.png}  &  \includegraphics[width=0.03\textwidth]{figures/figure_icon/cross.png}  &  \includegraphics[width=0.03\textwidth]{figures/figure_icon/check.png}  &  \includegraphics[width=0.03\textwidth]{figures/figure_icon/cross.png}  &  \includegraphics[width=0.03\textwidth]{figures/figure_icon/cross.png}  &  \includegraphics[width=0.03\textwidth]{figures/figure_icon/cross.png}  &  \includegraphics[width=0.03\textwidth]{figures/figure_icon/cross.png}  &  \includegraphics[width=0.03\textwidth]{figures/figure_icon/cross.png}  &  \includegraphics[width=0.03\textwidth]{figures/figure_icon/cross.png}  &  \includegraphics[width=0.03\textwidth]{figures/figure_icon/cross.png}  \\

\hline

% %%%%%%%%%%%%% Usability / Interactivity %%%%%%%%%%%%%
\multirow{7}{*}{\rotatebox[origin=c]{90}{Usability / Interactivity}}

% Is installation available across platforms?
& cross-platform                &  \includegraphics[width=0.03\textwidth]{figures/figure_icon/check.png}  &  \includegraphics[width=0.03\textwidth]{figures/figure_icon/check.png}  &  \includegraphics[width=0.03\textwidth]{figures/figure_icon/check.png}  &  \includegraphics[width=0.03\textwidth]{figures/figure_icon/check.png}  &  \includegraphics[width=0.03\textwidth]{figures/figure_icon/check.png}   &  \includegraphics[width=0.03\textwidth]{figures/figure_icon/cross.png}   &  \includegraphics[width=0.03\textwidth]{figures/figure_icon/cross.png}  &  \includegraphics[width=0.03\textwidth]{figures/figure_icon/check.png}   &  \includegraphics[width=0.03\textwidth]{figures/figure_icon/check.png}   &  \includegraphics[width=0.03\textwidth]{figures/figure_icon/check.png}  &  \includegraphics[width=0.03\textwidth]{figures/figure_icon/check.png}  &  \includegraphics[width=0.03\textwidth]{figures/figure_icon/cross.png}  \\

% is special privilege required to install the software?
& no system privilege required       &  \includegraphics[width=0.03\textwidth]{figures/figure_icon/check.png}  &  \includegraphics[width=0.03\textwidth]{figures/figure_icon/check.png}  &  \includegraphics[width=0.03\textwidth]{figures/figure_icon/check.png}  &  \includegraphics[width=0.03\textwidth]{figures/figure_icon/check.png}  &  \includegraphics[width=0.03\textwidth]{figures/figure_icon/check.png}  &  \includegraphics[width=0.03\textwidth]{figures/figure_icon/cross.png}  &  \includegraphics[width=0.03\textwidth]{figures/figure_icon/check.png}  &  \includegraphics[width=0.03\textwidth]{figures/figure_icon/check.png}  &  \includegraphics[width=0.03\textwidth]{figures/figure_icon/check.png}  &  \includegraphics[width=0.03\textwidth]{figures/figure_icon/check.png}  &  \includegraphics[width=0.03\textwidth]{figures/figure_icon/cross.png}  &  \includegraphics[width=0.03\textwidth]{figures/figure_icon/cross.png}   \\

% Can users rely on an interface?
& GUI available      &  \includegraphics[width=0.03\textwidth]{figures/figure_icon/check.png}  &  \includegraphics[width=0.03\textwidth]{figures/figure_icon/cross.png}   &  \includegraphics[width=0.03\textwidth]{figures/figure_icon/cross.png}  &  \includegraphics[width=0.03\textwidth]{figures/figure_icon/cross.png}  &  \includegraphics[width=0.03\textwidth]{figures/figure_icon/cross.png}  &  \includegraphics[width=0.03\textwidth]{figures/figure_icon/cross.png}   &  \includegraphics[width=0.03\textwidth]{figures/figure_icon/cross.png}  &  \includegraphics[width=0.03\textwidth]{figures/figure_icon/cross.png}  &  \includegraphics[width=0.03\textwidth]{figures/figure_icon/check.png}  &  \includegraphics[width=0.03\textwidth]{figures/figure_icon/check.png}  &  \includegraphics[width=0.03\textwidth]{figures/figure_icon/check.png}  &  \includegraphics[width=0.03\textwidth]{figures/figure_icon/cross.png}   \\

% is the tool able to deal with all available coding languages?
& language-agnostic \textsuperscript{*}            &  \includegraphics[width=0.03\textwidth]{figures/figure_icon/cross.png}   &  \includegraphics[width=0.03\textwidth]{figures/figure_icon/cross.png}   &  \includegraphics[width=0.03\textwidth]{figures/figure_icon/cross.png}   &  \includegraphics[width=0.03\textwidth]{figures/figure_icon/cross.png}  &  \includegraphics[width=0.03\textwidth]{figures/figure_icon/cross.png}  &  \includegraphics[width=0.03\textwidth]{figures/figure_icon/cross.png}  &  \includegraphics[width=0.03\textwidth]{figures/figure_icon/check.png}  &  \includegraphics[width=0.03\textwidth]{figures/figure_icon/check.png}  &  \includegraphics[width=0.03\textwidth]{figures/figure_icon/check.png}  &  \includegraphics[width=0.03\textwidth]{figures/figure_icon/check.png}  &  \includegraphics[width=0.03\textwidth]{figures/figure_icon/check.png}  &  \includegraphics[width=0.03\textwidth]{figures/figure_icon/check.png}  \\

% Can users extend the tool with custom modules/plugins?
& plugin possibility            &  \includegraphics[width=0.03\textwidth]{figures/figure_icon/check.png}  &  \includegraphics[width=0.03\textwidth]{figures/figure_icon/cross.png}  &  \includegraphics[width=0.03\textwidth]{figures/figure_icon/cross.png}  &  \includegraphics[width=0.03\textwidth]{figures/figure_icon/cross.png}  &  \includegraphics[width=0.03\textwidth]{figures/figure_icon/check.png}  &  \includegraphics[width=0.03\textwidth]{figures/figure_icon/check.png}   &  \includegraphics[width=0.03\textwidth]{figures/figure_icon/check.png}  &  \includegraphics[width=0.03\textwidth]{figures/figure_icon/check.png}   &  \includegraphics[width=0.03\textwidth]{figures/figure_icon/check.png}   &  \includegraphics[width=0.03\textwidth]{figures/figure_icon/check.png}  &  \includegraphics[width=0.03\textwidth]{figures/figure_icon/check.png}  &  \includegraphics[width=0.03\textwidth]{figures/figure_icon/check.png}   \\

% Can the tool run on HPC or cloud out of the box?
& HPC/cloud ready \textsuperscript{\ddag}              &  \includegraphics[width=0.03\textwidth]{figures/figure_icon/cross.png}  &  \includegraphics[width=0.03\textwidth]{figures/figure_icon/cross.png}  &  \includegraphics[width=0.03\textwidth]{figures/figure_icon/cross.png}  &  \includegraphics[width=0.03\textwidth]{figures/figure_icon/cross.png}  &  \includegraphics[width=0.03\textwidth]{figures/figure_icon/cross.png}  &  \includegraphics[width=0.03\textwidth]{figures/figure_icon/check.png}  &  \includegraphics[width=0.03\textwidth]{figures/figure_icon/check.png}  &  \includegraphics[width=0.03\textwidth]{figures/figure_icon/check.png}  &  \includegraphics[width=0.03\textwidth]{figures/figure_icon/cross.png}  &  \includegraphics[width=0.03\textwidth]{figures/figure_icon/check.png}  &  \includegraphics[width=0.03\textwidth]{figures/figure_icon/cross.png}  &  \includegraphics[width=0.03\textwidth]{figures/figure_icon/check.png}  \\

% Can the tool support or execute GUI-based scientific applications (e.g., PyQt, Tkinter, Shiny, Jupyter, etc.) that are part of the user's environment or workflow?
& Supports GUI application execution \textsuperscript{\S} &  \includegraphics[width=0.03\textwidth]{figures/figure_icon/check.png}  &  \includegraphics[width=0.03\textwidth]{figures/figure_icon/check.png}  &  \includegraphics[width=0.03\textwidth]{figures/figure_icon/check.png}  &  \includegraphics[width=0.03\textwidth]{figures/figure_icon/check.png}  &  \includegraphics[width=0.03\textwidth]{figures/figure_icon/check.png}  &  \includegraphics[width=0.03\textwidth]{figures/figure_icon/cross.png}  &  \includegraphics[width=0.03\textwidth]{figures/figure_icon/cross.png}  &  \includegraphics[width=0.03\textwidth]{figures/figure_icon/cross.png}  &  \includegraphics[width=0.03\textwidth]{figures/figure_icon/check.png}  &  \includegraphics[width=0.03\textwidth]{figures/figure_icon/cross.png}  &  \includegraphics[width=0.03\textwidth]{figures/figure_icon/cross.png}  &  \includegraphics[width=0.03\textwidth]{figures/figure_icon/cross.png} \\

\hline

% clarification
\multicolumn{14}{c}{\makecell[l]{*\quad Refers to whether the tool can support or execute steps in any language, not necessarily the language it's implemented in.}} \\
\multicolumn{14}{c}{\makecell[l]{\dag \quad Refers to the ability to natively generate a Docker/Apptainer image or definition file from the tool’s output.}} \\
\multicolumn{14}{c}{\makecell[l]{\ddag \quad Means the tool has native or first-party support/integration with job schedulers (e.g., SLURM) or cloud orchestration (e.g., Kubernetes), not merely that it can be run manually on those systems.}} \\
\multicolumn{14}{c}{\makecell[l]{\P \quad Tools marked with exclamation mark can expose dependencies only under certain configurations (e.g., if not using opaque containers).}} \\
\multicolumn{13}{c}{\makecell[l]{\S \quad Indicates whether GUI-based applications (e.g., PyQt, Tkinter, Shiny) can be executed from within the tool’s environment or workflows. Exclamation marks cases where additional system configuration (e.g., X11 forwarding, desktop integration, or container options) is required.}} \\

% legend
\multicolumn{14}{c}{\makecell[l]{  \includegraphics[width=0.03\textwidth]{figures/figure_icon/check.png}  \quad Feature is fully supported or clearly true. }} \\
\multicolumn{14}{c}{\makecell[l]{  \includegraphics[width=0.03\textwidth]{figures/figure_icon/cross.png}  \quad Feature is partially supported, conditionally true, or depends on usage context. }} \\
\multicolumn{14}{c}{\makecell[l]{  \includegraphics[width=0.03\textwidth]{figures/figure_icon/cross.png}   \quad Feature is not supported or not applicable.}} \\

\end{tabular} 

}

\end{adjustbox}

\caption{Comparison of our framework to other tools, package managers, workflow engines and container applications}
\label{tab:tool_comparison}

\end{table*}
\end{landscape}

\clearpage
\newpage

\begin{table*}[ht]
    \centering
    \begin{tabular}{p{2cm}|p{2cm}|p{5cm}|p{3cm}|p{3cm}}
        Approach & illustrative coordination actions & Artifact integration components (beyond host adapter) & Knowledge needed & Primary change surface \\
        \hline
        \hline
        Album & 6 & backend identifiers, optional config & album API & solutions, boundary contract \\
        manual multi-environment & 10 & environment paths, optional config & Environment setup, subprocess API & environment, scripts, invocation logic, boundary contract \\
        Container & 6 & image tags, runtime flags, optional config & container API, runtime configuration & image, runtime configuration,  boundary contract\\
    \end{tabular}
    \caption{Work comparison table: The table shows a direct comparison between the approaches to reach the same goal of integrating two conflicting backends into the host python tool. Please note that counts are approximate and depend on how actions are grouped; they reflect coordination surfaces rather than runtime complexity.}
    \label{worktable}
\end{table*}

\clearpage
\newpage

\begin{table*}
    \centering
    \begin{tabular}{p{3.5cm}|p{12cm}}
    
    \multicolumn{2}{p{13cm}}{\textbf{Core Album execution concepts}} \\ \hline
    solution &
    A user-authored artifact (typically a single Python entry-point file; optionally a directory) that Album evaluates to obtain a manifest and lifecycle hooks implementing a scientific routine. \\

    manifest &
    A machine-readable description of a solution (metadata, parameters/arguments, and references to environment and runner requirements) that is validated against a JSON schema and used by Album to install and execute the routine reproducibly. \\

    lifecycle hook &
    A standardized entry function implementing a phase of a solution’s operation (e.g., install/setup, run, test). Album dispatches the appropriate hook depending on the user’s action. \\

    controller &
    The framework component running in the Album (host) environment that evaluates a solution to obtain its manifest, validates/migrates it, builds the per-solution environment, and delegates execution. \\

    runner &
    The minimal runtime component installed into the per-solution environment (e.g., the album-solution-api package) that reconstructs the manifest in that environment and dispatches the requested lifecycle hook. \\

    schema migration &
    The process of converting metadata/manifests authored against older schema versions into the latest schema version supported by the current Album installation, to preserve backward compatibility. \\
    
    \multicolumn{2}{p{13cm}}{\textbf{Dependency and environment management}} \\ \hline
    YAML, yaml or yml &
    A human-readable data serialization format commonly used for configuration. In Album, it is used to express environment specifications and dependency lists in a structured, machine-parseable way. \\

    package manager &
    Software that installs, updates, and removes packages, typically resolving dependencies from one or more repositories/registries (e.g., conda channels, PyPI). \\

    micromamba &
    A lightweight, conda-compatible package manager distributed as a standalone executable. In Album, it is used to create and install per-solution environments from environment specifications. \\

    conda channel &
    A named package repository in the conda ecosystem (e.g., conda-forge, Bioconda) used as a source for dependency resolution and installation. \\

    PyPI / pip &
    PyPI is the Python Package Index (a registry of Python packages). pip is the standard installer that retrieves and installs packages from PyPI (and compatible indexes). \\

    lock file (conda-lock) &
    A file that pins an environment to exact package versions (and typically platform-specific builds), improving repeatability of installations across time and machines. \\

    environment &
    An isolated runtime context containing a specific set of installed dependencies, often including an interpreter, libraries, and non-Python binaries, needed to run software in a controlled manner. In Album, environments are created per solution from an environment specification. \\

    \end{tabular}
    \caption{Glossary}
    \label{tab:glossary}
\end{table*}

\clearpage
\newpage

\setcounter{table}{4}
\renewcommand{\thetable}{\arabic{table}}
\renewcommand{\tablename}{Supplementary Table Cont.} 

\begin{table*}
    \centering
    \begin{tabular}{p{3.5cm}|p{12cm}}

    \multicolumn{2}{p{13cm}}{\textbf{Digital library, discovery, and collaboration}} \\ \hline
    catalog &  A git repository that stores solutions and includes a database index (SQLite) to support structured search, discovery, and web presentation of solutions and their metadata. \\

    collection &
    A local registry that aggregates multiple catalogs for a given Album installation, enabling unified search, updates, and management of available solutions. \\

    SQLite &
    A lightweight, file-based relational database used for local storage and querying of structured metadata (e.g., catalog and collection indices). \\

    CI/CD &
    Automated workflows triggered by code changes. CI typically builds and tests changes; CD typically packages, releases, or deploys artifacts after CI succeeds. \\

    Merge Request (MR), Pull Request (PR) &
    A proposed set of changes to a git repository that can be reviewed, discussed, and tested (often via CI) before being merged into a target branch. \\

    \multicolumn{2}{p{13cm}}{\textbf{Archival and citation}} \\ \hline
    container &
    A packaged filesystem image that bundles software with its dependencies and can be executed via a container runtime (e.g., Docker), supporting portability and long-term re-execution. \\

    DOI &
    A persistent identifier used to reference a specific digital artifact (e.g., a snapshot of code or data) in a citable, stable way. \\

    Zenodo &
    An open repository that can archive research artifacts and mint DOIs for specific releases/snapshots. \\
    
       \multicolumn{2}{p{13cm}}{\textbf{Automation and orchestration}} \\ \hline
    Application Programming Interface (API) &
    A documented interface (functions/classes, command-line calls, or web endpoints) that specifies how one software component interacts with another, what inputs it accepts, what outputs it returns, and what behavior/error conditions to expect. \\

    Model Context Protocol (MCP) &
    A protocol that lets an LLM access external tools through a standardized tool-calling interface (e.g., to discover, install, and execute Album solutions). \\

    Large Language Model (LLM) &
    A machine-learning model trained to generate and transform natural language (and often code). In this work, LLMs can generate Album solutions and orchestrate their execution via MCP. \\

    \end{tabular}
    \caption{Glossary}
\end{table*}

\includegraphics[width=0pt,height=0pt]{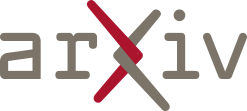}

\end{document}